\documentclass[reprint,prb,aps,twocolumn,superscriptaddress,longbibliography,10pt]{revtex4-1}
\usepackage{amsmath}
\usepackage[english]{babel}
\usepackage{graphicx}
\usepackage{color}
\usepackage{sidecap}
\usepackage[dvipsnames]{xcolor}

\usepackage[colorlinks,
            citecolor = blue, 
						linkcolor = blue,
						urlcolor  = blue,
						anchorcolor = blue,
						breaklinks = true
						]{hyperref}

\usepackage{soul}
\usepackage{amssymb}
\usepackage{bm}
\newif\ifdraft
\drafttrue
\sethlcolor{orange}
\usepackage[normalem]{ulem}
\newcommand{\addMisha}[1]{\textcolor{blue}{#1}}

\begin{document}

% ----- Title -----

\title{Long-range ferroelectric order in two dimensional excitonic insulators}

\author{M. M. Glazov}
\affiliation{Ioffe Institute, 194021 St. Petersburg, Russia}

\author{A. \.{I}mamo\u{g}lu}
\affiliation{Institute for Quantum Electronics, ETH Zürich, CH-8093 Zürich, Switzerland}

\date{\today}% It is always \today, today,

\begin{abstract} 
It is generally argued that Mermin-Wagner theorem excludes the possibility of long-range order in two dimensional bosonic systems at non-zero temperatures. In contrast, we show here that generic bilayer semiconductors could demonstrate true Bose-Einstein condensation of interlayer excitons. We show that the key requirements include (i) reduction of the interlayer band gap using an applied electric field so that excitons spontaneously appear in the ground state, (ii) band structure that allows for long-range electron-hole exchange interaction, and (iii) a finite magnetic field. Our results indicate that superfluidity and ferroelectric order can co-exist in two dimensional excitonic insulators.
\end{abstract}

\maketitle

% ----- Show line numbers -----
% Please comment out \linenumbers to disable line numbers
%\linenumbers

\section{Introduction and motivation}

\begin{figure}[b]
    \centering
    \includegraphics[width=0.75\linewidth]{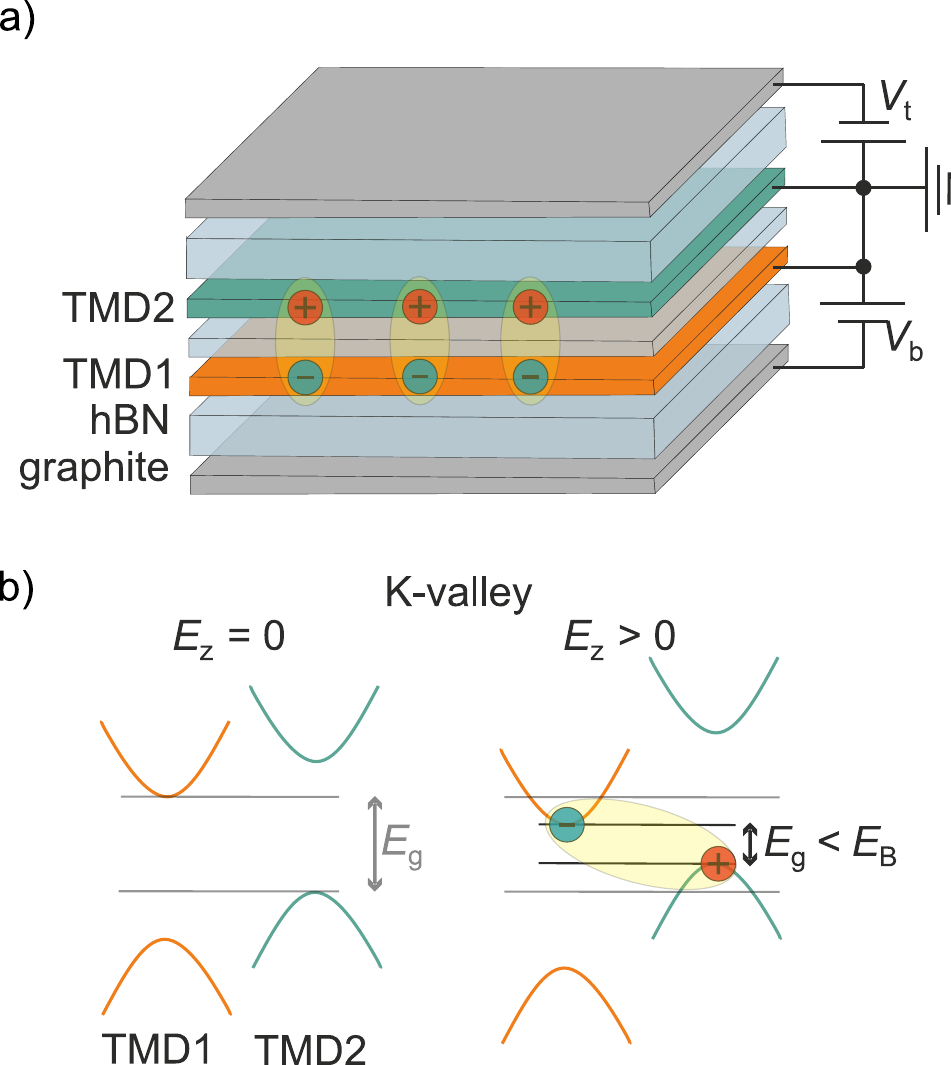}
    \caption{(a) The van der Waals heterostructure that we analyze. The two Transition Metal Dichalcogenide (TMD) layers have type II band alignment and they are either in direct contact or are separated by monolayer hexagonal Boron Nitride (hBN). Voltages applied on top and bottom graphite gates are used to adjust the vertical electric field $E_z$. (b) Sketch of the band diagram around the K-valley. The spatially indirect band gap, $E_g$ can be tuned via electric field $E_z$ to reach $E_g < E_B$ to ensure that interlayer excitons appear spontaneously. }
    \label{fig:scheme}
\end{figure}

Interlayer excitons in Transition Metal Dichalcogenide (TMD) bilayers constitute a promising platform for observation of the elusive excitonic insulator state.\cite{keldysh1965possible,PhysRev.158.462,PhysRevLett.99.146403,PhysRevB.90.155116,Du:2017aa,Kogar:2017aa,Ataei:2021aa,PhysRevLett.133.217002} By applying a finite electric field (Fig.~\ref{fig:scheme}a), it is possible to reduce the interlayer band-gap $E_g$, defined as the energy difference between the conduction band minimum of one layer and the valence band maximum of the other layer, so that it is smaller than the binding energy of the interlayer 1s exciton ($E_B$): in this limit interlayer excitons appear spontaneously in the ground-state.\cite{Ma:2021aa} However, Mermin-Wagner theorem excludes the possibility for continuous symmetry breaking at finite temperature $T$ for two dimensional (2D) systems with short range hopping or interactions.\cite{PhysRevLett.17.1133,PhysRev.158.383} Because the center-of-mass motion of interlayer excitons generically have a parabolic dispersion that is linked to short-range hopping, it is widely assumed that a $T > 0$ excitonic insulator state with true long-range order is not possible even in the limit $E_B > E_g$.

On the other hand, it is well known that coupling of optically active 2D intralayer excitons to the three-dimensional electromagnetic field vacuum substantially modifies their excitation spectra: within the light cone, light-matter coupling leads to a finite spontaneous emission rate. Outside the light cone, the strong interaction with electromagnetic near field results in a splitting of the excitonic bands: while excitons that are polarized transversally (T) to the center-of-mass momentum $k$ exhibit parabolic dispersion, longitudinally polarized (L) excitons acquire linear dispersion.\cite{PhysRevB.41.7536,glazov2014exciton,Yu:2014fk-1,Qiu2015} The origin of linear dispersion can be traced back to flip-flop dipole-dipole interaction that leads to long-range hopping, scaling as $1/r^3$.\cite{agranovich:galanin,PhysRevB.111.045301} However, since the contribution of light-matter coupling to exciton self-energy is purely imaginary for $k \rightarrow 0$, finite energy intralayer excitons at $E_x(k=0) = E_x = E_g - E_B > 0$ cannot exhibit continuous symmetry breaking, even though a laser-like nonequilibrium transition to a coherent state with weak phase diffusion cannot be ruled out.

In this work, we show that for ground state interlayer excitons under external magnetic field, the exciton dispersion in the limit $k \rightarrow 0$ is linear, provided that $k \neq 0$ excitons have in-plane dipolar coupling to the electromagnetic field. After carrying out a mean field analysis using a BCS variational Ansatz to show the formation of a condensate for $T=0$ without any external phase fixing, we analyze the spectrum of excitations and derive how coupling to the electromagnetic field modifies the dispersion of the elementary Bogoluibov excitations. Finally, we show that the thermally excited exciton density remains finite for $T > 0$, signaling the emergence of a true long-range order at a critical temperature $T_c > 0$. 

Our analysis assumes that the interlayer excitons in the limit $E_g > E_B$ occupy a single valley and couple either to right- or left-hand circularly polarized fields within the light cone. When $E_B > E_g$ and $T >T_c$, the excitons do not have a well defined in-plane polarization but a finite small static polarization along the direction $z$ that separates the two layers. Upon condensation, the ground-state excitons acquire a large in-plane polarization, realizing a ferroelectric phase transition that is not linked to a structural phase transition. We emphasize that the emergence of an ordered ferroelectric phase upon spontaneous breaking of $U(1)$ symmetry associated with in-plane exciton polarization constitutes an example where vacuum fluctuations qualitatively modify the ground state of an electronic system.

%Here we provide a brief description of Mermin-Wagner, focusing on how one can understand the physics behind its statement. We then mention that particles with long range hopping could have sub-quadratic dispersion which in turn allows for true BEC phase transition. 

%We then describe excitons and their dispersion which at first sight does not look promising due to imaginary self-energy ensuring lack of dispersion for momenta smaller than the momentum of light at energy = exciton energy. We then state that this begs the question if dipolar excitons in the ground state could have linear dispersion extending all the way to k=0.

\section{Derivation of the exciton gap equation}

Here we show the steps leading to the exciton gap equation, identifying the linear and nonlinear contributions arising from electron-hole attraction, as well as the effects of the electron-electron and hole-hole interactions and phase space filling. We then show that the gap equation admits a nontrivial solution indicating excitonic instability if the band-gap is smaller than the exciton binding energy.

What distinguishes our analysis from prior work is the inclusion of inter-layer coupling: for direct tunnel coupling between the conduction and valence bands, we obtain a source term for the order parameter that fixes the phase and amplitude; the resulting gap equation is reminiscent of resonant laser excitation of excitons. Remarkably, when the interband coupling is of chiral $p$-wave form, that is, it is proportional to $k_x + ik_y$, no phase fixing occurs. As we further show, inter-layer coupling of this form, together with (a) exciton binding energy larger than the bare band-gap, and (b) application of the magnetic $B$-field, leads to linear dispersion of excitons as $k \to 0$, which in turn ensures the existence of true long-range order. 

\subsection{Hamiltonian of the bilayer system}

We consider a heterobilayer structure with type-II band alignment depicted in Fig.~\ref{fig:scheme}. Let layer 1 be the ``electron'' layer and layer 2 be the ``hole'' layer, i.e., the conduction band minimum of the bilayer corresponds to the layer 1 and the valence band minimum to the layer 2. We disregard the spin and valley degrees of freedom considering the bottom conduction band and top valence band in one of the valleys $K_+$ or $K_-$\addMisha{.}
%; we will analyze the impact of two valleys below, see also \addMisha{Appendix AAA}. \commentMisha{We probably need to add a schematic picture with the studied system}

We assume that the tunneling is allowed between the two conduction (valence) bands of the two layers but that there is no interband tunneling. Consequently, the interlayer hybridization results from the valence band 1 -- valence band 2 and conduction band 1 -- conduction band 2 tunnel couplings. The $4\times 4$ Hamiltonian describing orbital states (without Coulomb interaction) reads
\begin{equation}
\mathcal H_4 = 
\begin{pmatrix}
E_{c1}(\bm k) & \gamma^1_\alpha k_\alpha  & t_{cc} & 0 \\
\gamma^{1,*}_\alpha k_\alpha & E_{v1}(\bm k) & 0 & t_{vv} \\
t_{cc}^* & 0 & E_{c2}(\bm k) & \gamma^{2,*}_\beta k_\beta \\
0 & t_{vv}^* & \gamma^2_\beta k_\beta & E_{v2}(\bm k)
\end{pmatrix}.
\end{equation}
Here the $2\times 2$ blocks correspond to the first and second layers, $E_{c1}(\bm k) = E_{c1} + \hbar^2 k^2/2m_c$ ($E_{c2}(\bm k) = E_{c2} + \hbar^2 k^2/2m_c$) and $E_{v1}(\bm k) = E_{v1} + \hbar^2 k^2/2m_v$ ($E_{v2}(\bm k) = E_{v2} + \hbar^2 k^2/2m_v$) are the ``bare'' electron dispersions in the conduction and valence bands of the first (second) layer, respectively. $E_{c1}(\bm k)$ ($E_{c2}(\bm k)$) and $E_{v1}(\bm k)$ ($E_{v2}(\bm k)$) are the conduction and valence band energies with the effective masses $m_c$ and $m_v$ determined by couplings to other remote bands that are not described by the Hamiltonian of Eq.(1). The parameters $t_{cc}$ and $t_{vv}$ describe the interlayer tunneling strength and $\gamma^1_\alpha k_\alpha$, $\gamma^2_\beta k_\beta$ describe the $\bm k\cdot \bm p$-mixing of the conduction and valence bands in the corresponding layers with $\alpha,\beta = x,y$ denoting the cartesian components, and $\gamma^1_\alpha$, $\gamma^2_\beta$ are the interband momentum matrix elements multiplied by $\hbar/m_0$ ($m_0$ is the free electron mass). Hereafter the summation over the repeated subscripts is implicitly assumed.  For $K_\pm$ valleys $\gamma^{1,2}_\alpha k_\alpha = \hbar p^{1,2}_{cv}/m_0(k_x \mp \mathrm i k_y)$. The interband matrix elements $p_{cv}^{1,2}$ can be made real by the choice of the wavefunction phases.

Assuming that in the relevant wavevector range $E_{ci} - E_{vi} \gg |\gamma_\alpha^i k_\alpha|$ ($i=1,2$), $|E_{c1} - E_{c2}| \gg |t_{cc}|$, and $|E_{v1} - E_{v2}| \gg |t_{vv}|$, we use perturbation theory to reduce the full Hamiltonian to the $2\times 2$ matrix describing the nearest conduction $c1$ and valence $v2$ band states:\cite{birpikus_eng}
\begin{equation}
\label{H2}
\mathcal H_2 = 
\begin{pmatrix}
E_{c}(\bm k)  & \gamma_\alpha k_\alpha\\
\gamma_\alpha^* k_\alpha & E_{v}(\bm k)
\end{pmatrix},
\end{equation}
where
\begin{subequations}
\begin{align}
E_c(\bm k) = E_{c} + \frac{\hbar^2 k^2}{2m_c} + \frac{|t_{cc}|^2}{E_{c1} - E_{c2}} + \frac{|\gamma^1_\alpha k_\alpha|^2}{E_{c1} - E_{v1}} \nonumber \\
= \frac{E_g}{2} + \frac{\hbar^2 k^2}{2m_e}, \\
E_v(\bm k) = E_{v} + \frac{\hbar^2 k^2}{2m_v} + \frac{|t_{vv}|^2}{E_{v2} - E_{v1}} + \frac{|\gamma^2_\alpha k_\alpha|^2}{E_{v2} - E_{c2}} \nonumber \\= - \frac{E_g}{2} - \frac{\hbar^2 k^2}{2m_h}, \\
\gamma_\alpha k_\alpha = %\underbrace{
\frac{\gamma^1_\alpha k_\alpha t_{vv}}{2} \left(\frac{1}{E_{c1} - E_{v1}}+\frac{1}{E_{v2} - E_{v1}}\right)
%}_{c1\leftrightarrow v1\leftrightarrow v2}   
\nonumber\\ +
%\underbrace{
\frac{\gamma^{2,*}_\alpha k_\alpha t_{cc}}{2} \left(\frac{1}{E_{c1} - E_{c2}}+\frac{1}{E_{v2} - E_{c2}}\right).
%}_{c1\leftrightarrow c2\leftrightarrow v2}.
\end{align}
\end{subequations}
Here $m_e$, $m_h>0$ are the effective masses of the electrons and holes in the absence of interlayer coupling, $E_g$ is the effective band gap, $\gamma_\alpha$ are the effective interband momentum matrix elements. To proceed, we assume that $\gamma_\alpha$ are much smaller than the other relevant energy scales and the dispersion of the charge carriers is mainly controlled by the diagonal terms in the Hamiltonian provided $E_g \gg |\gamma_\alpha k_\alpha|$.

 We rewrite the Hamiltonian~\eqref{H2} in the second quantization representation introducing the conduction band creation (annihilation) operators $a_{\bm k}^\dag$ ($a_{\bm k}$) and the valence band hole creation (annihilation) operators $b_{\bm k}^\dag = a_{v,-\bm k}$ ($b_{\bm k} = a_{v,-\bm k}^\dag$). Equation~\eqref{H2} reads
\begin{multline}
\label{H2:2:nonint}
\mathcal H_2 = \sum_{\bm k} E_c (\bm k) a^\dag_{\bm k} a_{\bm k} + \sum_{\bm k} E_h (\bm k) b^\dag_{\bm k} b_{\bm k} \\ + \sum_{\bm k} \left( {\Lambda_{\bm k}}  a^\dag_{\bm k} b^\dag_{-\bm k} + {\rm h.c.}\right),
\end{multline}
where we introduced the hole dispersion $E_h(\bm k) = -E_v(-\bm k)$ and interband coupling $\Lambda_{\bm k} = \gamma_\alpha k_\alpha$. The coupling has a chiral $p$-wave form related to the different symmetries of the conduction and valence bands and allowed optical transition between the bands. If direct interband tunneling were possible, $\Lambda_{\bm k}$ would contain a wavevector-independent contribution.

Excitons form because of the Coulomb attraction obetween electrons and holes. Let $V_{\bm q}>0$ be the Fourier transform of the Coulomb interaction potential within the same layer and let $d$ be the interlayer distance.
%For simplicity we assume that its form the same for interaction between the quasiparticles in the same and different layers. 
Hence, the electron-electron, hole-hole, and  electron-hole contributions read
\begin{multline}
\label{Coulomb:dir}
\mathcal V = \frac{1}{2}\sum_{\bm k, \bm k', \bm q} V_{\bm q}\left(a_{\bm k}^\dag a_{\bm k'}^\dag a_{\bm k' + \bm q} a_{\bm k - \bm q} + b_{\bm k}^\dag b_{\bm k'}^\dag b_{\bm k' + \bm q} b_{\bm k - \bm q}\right. \\ \left.-2{e^{-qd}}a_{\bm k}^\dag b_{\bm k'}^\dag b_{\bm k' + \bm q} a_{\bm k - \bm q} \right).
\end{multline}
The prefactor $e^{-qd}$ takes into account slight reduction of the interlayer interaction as compared to the intralayer one, see Ref.~\onlinecite{Semina:2019aa} for more advanced models. 
Equation~\eqref{Coulomb:dir} takes into account the static Coulomb interaction which is responsible for formation of the excitons. We will analyze the long-range exchange interaction related to the macroscopic electromagnetic fields produced by the excitons in Sec.~III using an electrodynamical approach.

%\commentMisha{We can shorten this part omitting all the technical details and putting them to the supplement/appendix} \commentAtac{I agree to put the details in the supplementary. We could also directly write the exp[-qd]factor in electron-hole interaction} \commentMisha{I have added an exponential factor and a comment above. Feel free to revise this text as you prefer.}

\subsection{Gap equation and order parameter}

We now analyze the possibility of exciton insulator formation. It is convenient to assume the Bardeen-Cooper-Schrieffer (BCS) form of the manybody wavefunction
\begin{equation}
    \label{BCS}
    \Psi = \prod_{\bm k} \left(u_{\bm k} + v_{\bm k} a^\dag_{\bm k} b^\dag_{-\bm k}\right)|0\rangle,
\end{equation}
where $|0\rangle$ is the ground state of the system with the filled valence band and the empty conduction band and $u_{\bm k}, v_{\bm k}$ are the coefficients which together quantify electron-hole pairing. We consider the $T=0$ case and study the pairing in the lowest energy state corresponding to zero net momentum of excitons. Neglecting for simplicity the electron-electron and hole-hole repulsion and making standard transformations~\cite{tolmachev58,keldysh1965possible,KozlovMaksimov,kozlovmaksimov65,PhysRev.158.462,GuseinovKeldysh,lozovik:yudson,Lozovik:1977aa,Littlewood:2004aa,PhysRevB.107.075105,PhysRevB.106.125311} we obtain the exciton gap $\Delta_{\bm k} = \sum_{\bm k'} V_{\bm k - \bm k'} u_{\bm k'}^* v_{\bm k'}$ equation in the form
\begin{equation}
    \label{gap}
    \Delta_{\bm k} = \sum_{\bm k'} V_{\bm k- \bm k'} \frac{\Delta_{\bm k'}}{\mathcal E_{\bm k'}} + \Lambda_{\bm k},
\end{equation}
where the renormalized energy reads 
\begin{equation}
\label{renormalized}
    \mathcal E_{\bm k} =\sqrt{[E_e(\bm k)+E_h(-\bm k)]^2 + 4\Delta_{\bm k}^2}.
\end{equation}
We analyze the effect of  finite $T$ in Appendix~\ref{sec:gap:T}.

For low condensate density, $\Delta_{\bm k}\to 0$, Eq.~\eqref{gap} can be recast in somewhat different form convenient for the following analysis. To that end we decomponse $\mathcal E_{\bm k}$ into series in $\Delta_{\bm k}$ keeping only the zero and second-order terms and introduce the wavefunction $\Psi_{\bm k} = u_{\bm k}^* v_{\bm k} =\Delta_{\bm k}/\mathcal E_{\bm k} \approx \Delta_{\bm k}/ [E_e(\bm k)+E_h(-\bm k)] $ which, upon expanding $\Delta_{\bm k}/\mathcal E_{\bm k}$ to lowest order, gives:
\begin{multline}
\label{exciton:cond:full}
    [E_e(\bm k)+E_h(-\bm k)]\Psi_{\bm k} -\sum_{\bm k} V_{\bm k - \bm k'} \Psi_{\bm k'}\\ + 2\sum_{\bm k'} V_{\bm k- \bm k'}|\Psi_{\bm k'}|^2 \Psi_{\bm k'}=\Lambda_{\bm k}.
\end{multline}
We note that inclusion of the repulsion between the charge carriers of the same polarity modifies the nonlinear term in Eq.~\eqref{exciton:cond:full}, see Ref.~\onlinecite{GuseinovKeldysh} and Appendix~\ref{sec:gap:alternative}.

The nonlinear gap equation (Eq.~\eqref{gap} or Eq.~\eqref{exciton:cond:full}) admits, in general, different types of solutions. The first type correspond to solutions of inhomogeneous Eqs.~\eqref{gap}, \eqref{exciton:cond:full} that vanish as $\Lambda_{\bm k} \to 0$. These induced or forced solutions are somewhat trivial; since they correspond to the presence of interband correlations as a result of the coupling described by the term $\propto \Lambda_{\bm k}$ in the Hamiltonian~\eqref{H2:2:nonint}. Importantly, there is a second class of solutions of the gap equation which are non-zero as $\Lambda_{\bm k} \to 0$. These solutions of the second type, which describe spontaneous pairing in the system, are what we focus on.

\begin{figure}[tb]
    \centering
    \includegraphics[width=\linewidth]{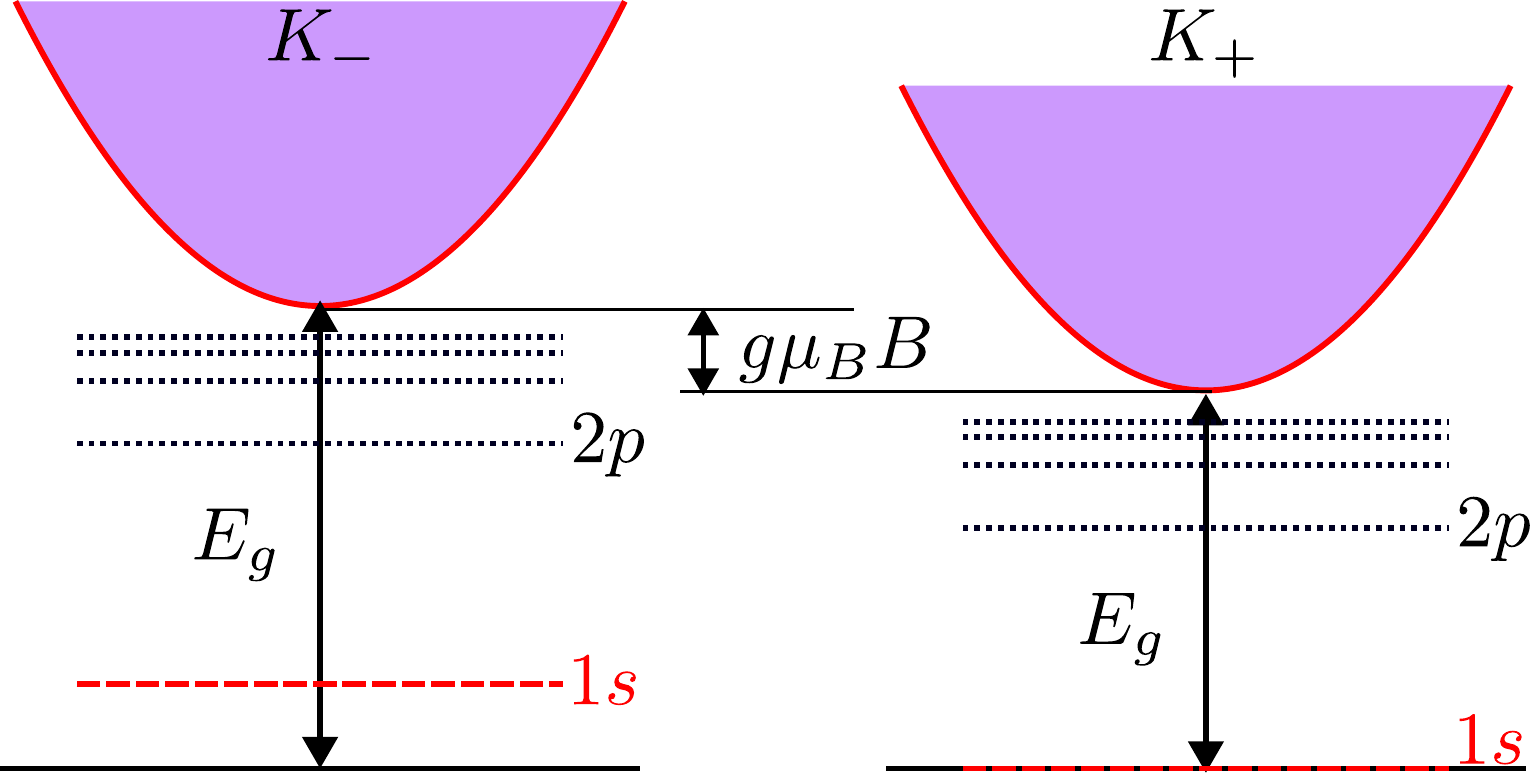}
    \caption{Schematic illustration of the heterobilayer excitonic band structure in the presence of magnetic field induced Zeeman splitting $g\mu_B B$, where $g$ is the exciton $g$-factor, $\mu_B$ is the Bohr magneton and $B$ is the normal component of the magnetic field. Series of bound excitonic states are shown by the dashed ($1s$ state) and dotted ($2p$ and other higher-lying states) lines. Parabolas show the continuum states of electron-hole pairs. It is seen that the exciton insulator instability condition~\eqref{cond:pairing} is first met for $1s$ exciton in the $K_+$ valley.}
    \label{fig:Bfield:tune}
\end{figure}

We first set $\Lambda_{\bm k}=0$ and analyze the instabilities in the system. In the vicinity of the pairing onset $\Psi_{\bm k}$ in Eq.~\eqref{exciton:cond:full} is vanishingly small and the nonlinear term becomes negligible. Thus, Eq.~\eqref{exciton:cond:full} reduces to the Schr\"odinger equation describing the Coulomb attraction of the electron and the hole which has nontrivial solutions provided that the Coulomb interaction is sufficiently large to ensure that the lowest energy bound eigenstate has zero energy, i.e., the binding energy of this state is equal to the  band gap $E_g$. Naturally, when reducing $E_g$ using the applied electric field, the first state to reach zero energy is the ground, $1s$ exciton state (Fig.~\ref{fig:Bfield:tune}). Hence, under the condition
\begin{equation}
    \label{cond:pairing}
    E_B \geqslant E_g,
\end{equation}
where $E_B$ is the $1s$ exciton binding energy, excitons form spontaneously in the ground state. The wave function can be written as
\begin{equation}
    \label{Psi:k}
    \Psi_{\bm k} = a \varphi^{1s}_{\bm k} = |a| e^{\mathrm i \chi} \varphi^{1s}_{\bm k},
\end{equation}
where $\varphi^{1s}_{\bm k}$ is the normalized $1s$ exciton wavefunction in momentum space, $a$ is the complex parameter with its absolute value $|a|$ and phase $\chi$.  The amplitude $a$ of the exciton condensate can be found from Eq.~\eqref{exciton:cond:full} by substituting $\Psi_{\bm k}$ in the form of Eq.~\eqref{Psi:k}, multiplying by ${\varphi^{1s}_{\bm k}}^*$ and summing over $\bm k$
\begin{equation}
    \label{c:simple}
    (E_g - E_B) a =-g|a|^2a + \lambda,
\end{equation}
where we restored the inhomogeneous term,
\begin{equation}
    \lambda=\sum_{\bm k} {\varphi^{1s}_{\bm k}}^* \Lambda_{\bm k}.
\end{equation}
Equation~\eqref{c:simple} contains the effective nonlinear ``interaction'' term with the interaction parameter
\begin{equation}
    \label{gint}
    g=\sum_{\bm k}|\varphi^{1s}_{\bm k}|^4 \left(E_B + \frac{\hbar^2 k^2}{2\mu}\right)>0. 
\end{equation}
The electron-electron and hole-hole repulsion contributions that we neglected change the expression for $g$ but do not significantly alter the analysis and conclusions of this work (see Appendix~\ref{sec:gap:alternative} for details).

In the experimentally relevant case of TMD heterobilayers, the conduction and valence bands have different symmetry and the interband coupling terms $\Lambda_{\bm k}$ are propotional to the combinations of the wavevector components $k_\alpha$, see Eq.~(3c). Since the $1s$ exciton wavefunction is axially symmetric, $\lambda \equiv 0$. Equation~\eqref{c:simple} is therefore homogeneous and the phase of $a$ is arbitrary despite the presence of interband coupling. This is the distinguishing feature of the system we analyze in comparison to the previously investigated models where the bands were essentially of the same symmetry leading to  $\lambda\ne 0$, and the condensate phase was fixed by the interband coupling;\cite{GuseinovKeldysh,lozovik:yudson,Lozovik:1977aa,Littlewood:2004aa} the latter case is similar to the case where a coherent state of excitons is generated by a resonant laser field. Hence, in contrast to the common knowledge,\cite{GuseinovKeldysh,Littlewood:2004aa} in two-dimensional semiconductors with different symmetries of conduction and valence bands the exciton condensation upon varying $E_g$ is a second order quantum phase  transition: the phase of the condensate is arbitrary. Note that for the $\bm k$-linear interband coupling the phase could be fixed for $p$-excitons with odd envelope functions. However, their binding energies are significantly smaller than for $1s$ state. Hence, an instability in the system occurs first for the $1s$ exciton, where the phase is arbitrary. These results also follow from more general Eq.~\eqref{gap}.

The solution of Eq.~\eqref{c:simple} in the relevant case of $\lambda=0$ is simple:
\begin{equation}
\label{c}
a=\begin{cases}
0, \quad E_g > E_B,\\
e^{\mathrm i \chi} g^{-1/2} \sqrt{E_B - E_g}, \quad E_g <E_B,
\end{cases}
\end{equation}
where $\chi$ is an arbitrary phase. Naturally, as soon as the condition~\eqref{cond:pairing} is met, the order parameter spontaneously appears in the system. In accordance with Eq.~\eqref{BCS} the manybody wavefunction of the system can be represented as a coherent state:
\begin{equation}
\label{coherent}
\Psi \propto |0\rangle + a|1\rangle + \frac{a^2}{\sqrt{2!}} |2\rangle +  \frac{a^3}{\sqrt{3!}} |3\rangle \ldots,
\end{equation}
where $|{1}\rangle = \sum_{\bm k} \varphi_{\bm k}^{1s} a_{\bm k}^\dag b_{-\bm k}^\dag|0\rangle$ is the state with a single exciton in the system; states with higher exciton number ($|2\rangle, |3\rangle$) can be expressed similarly.
%$|\addMisha{2\rangle =  \sum_{\bm k} \varphi_{\bm k}^{1s} a_{\bm k}^\dag b_{-\bm k}^\dag|1\rangle}$ is the state with two excitons, $|\addMisha{3}\rangle$ is the state with three excitons, etc.

\begin{figure}[tb]
    \centering
    \includegraphics[width=\linewidth]{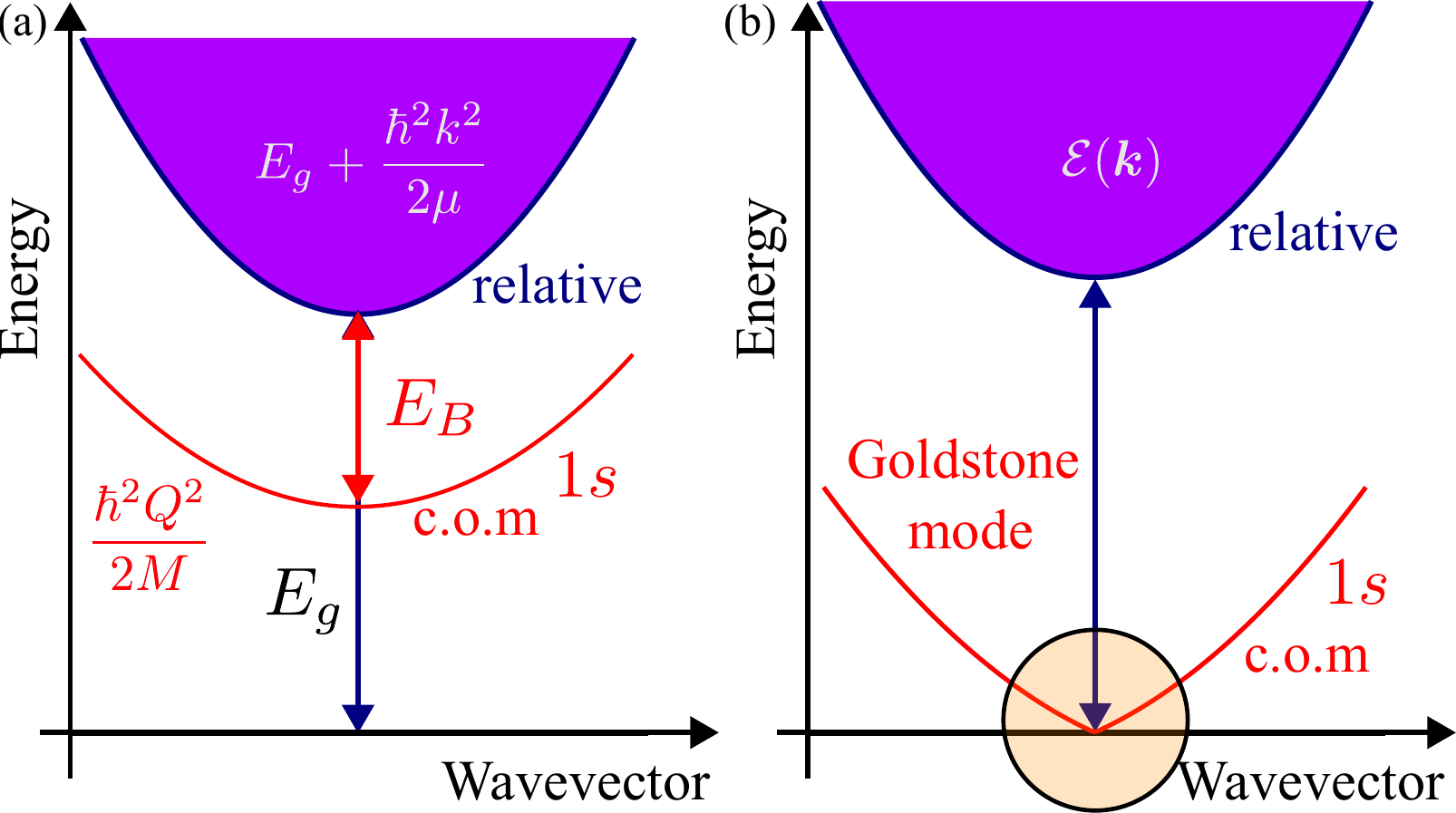}
    \caption{Spectrum of excitations. Only one bound state, $1s$, is shown. (a) Non-condensed system, $\Delta=0$. The excitations correspond to the exciton center of mass motion with parabolic dispersion (red parabola) and relative motion of electron-hole pairs in continuum (shaded region bounded by blue parabola). Here, we neglect the weak light-matter interaction. (b) Excitonic insulator state, $\Delta=0$. The relative motion continuum (shaded region bounded by blue parabola) is separated by the gap   $\sqrt{E_g^2+4\Delta^2}$ from the condensate ground state. Red curve sketches the dispersion of the collective, Goldstone, mode. Orange region of small momenta is of interest for analysis of the condensate stability.}
    \label{fig:excit:simple}
\end{figure}

So far we have analyzed the $T=0$ limit.  We now focus on the case where $E_B \ge E_g$, and consider the $T$ dependence of ground-state excitons. First temperature scale of relevance is the pairing temperature $T_{\rm pair}$  below which a thermal state of excitons appear: We estimate $T_{\rm pair}$ in Appendix~\ref{sec:gap:T}. Superfluidity without long-range order emerges for $T < T_{BKT}$, where $T_{BKT}$ is the Berezinskii-Kosterlitz-Thouless transition temperature~\cite{1971JETP...32..493B,1972JETP...34..610B,1973JPhC....6.1181K}. The critical temperature $T_c$ for condensation is lower and is linked to the suppression of phase fluctuations of the order parameter.  Since excitons are charge neutral, the lowest energy collective excitations correspond to the gapless Goldstone mode [Fig.~\ref{fig:excit:simple}(b)]. To analyze the stability of 
%\addMisha{The analysis above corresponds to the zero-temperature case. Allowance for the finite temperature via quasiparticle distribution in the gap equation~\eqref{gap} performed in Appendix~\ref{sec:gap:T} shows that the critical temperature for the pairing is high, $k_B T_c \sim E_B/\ln[E_B/(E_B-E_g)]$ at $E_B-E_g \ll E_B$, see Eq.~\eqref{Tc:log}. It is because such approach corresponds to the accounting for electron-hole pairing and unpairing processes only, they require energy $\sim \sqrt{E_g^2+4\Delta^2}$, see Fig.~\ref{fig:excit:simple}.}
%it disregards fluctuations of the order parameter itself -- the collective, Goldstone mode, similarly to the case of BCS scenario in the saddle point approximation.\cite{PhysRevB.55.15153,Altland_Simons_2010} The latter represent the lowest-energy excitations that are gapless, see Fig.~\ref{fig:excit:simple}(b). Hence, to analyze the stability of 
the exciton condensate we need to determine the spectrum of collective excitations, which, as shown below, are strongly affected by the coupling to the electromagnetic field.
%Hence, in the next section we evaluate the response of the excitonic insulator to the electromagnetic field and in Sec.~\ref{phase} we address the dispersion of collective mode related to the phase fluctuations and demonstrate the possibility of true condensation.

\section{Light-matter interaction}\label{lm}
In this Section, we analyze the response of the excitonic insulator to the electromagnetic field vacuum. We focus on the case where an external magnetic field ensures that the lowest energy interlayer exciton state is the $1s$ exciton in $K+$ valley due to valley Zeeman effect (Fig.~\ref{fig:Bfield:tune}). We also assume that the inter-valley states where the electron and hole occupy different valleys also have higher energy. In this regime, the pairing threshold~\eqref{cond:pairing} is reached in the $K_+$ valley where $\Lambda_{\bm k} \propto k_x - \mathrm i k_y$. To describe the light-matter interaction, we introduce the matrix element of the in-plane dipole moment of the interband exciton $\hat{\bm d}$
\begin{equation}
    \label{inter:dipole}
    \langle {\rm exc}| \hat{\bm d}|0\rangle = \frac{d_{\rm exc}}{\sqrt{2}} (\hat{\bm x} - \mathrm i\hat{\bm y}),
\end{equation}
where $\hat{\bm x}$, $\hat{\bm y}$ are the unit vectors along the respective coordinate axes and $d_{\rm exc} \propto  |\lambda_{\alpha}| \varphi_{1s}({\bf r} = 0)$  is a real parameter. For the exciton in the $K_-$ valley, the sign on the right hand side of Eq.~\eqref{inter:dipole} should be reversed. 
The exciton insulator described by the coherent state~\eqref{coherent} is characterized by a non-zero static dipole moment $\bm P$ with the in-plane components
\begin{equation}
    \label{P:static}
    P_x = \frac{d_{\rm exc}}{\sqrt{2}} (a+a^*), \quad   P_y =\frac{d_{\rm exc}}{\sqrt{2}}\mathrm i  (a-a^*),
\end{equation}
Hence, the polarization arises as a linear function of the condensate order parameter $a$ and its orientation in the heterostructure plane is determined by the phase $\chi$. Consequently, fixing of $\chi$ upon transition to the exciton insulator state can be viewed as a transition to a ferroelectric state with finite in-plane polarization. We remark that, because of the heterobilayer asymmetry, there is also a $z$-component of the static dipole moment
\begin{equation}
    \label{P:static:z}
    P_z = d_z |a|^2,
\end{equation}
which is quadratic in order parameter and controlled by the static $z$-component of the exciton dipole. In what follows, we assume the limit of small $|a|$ and ignore $P_z \propto |a|^2$ in comparison to $P_x,P_y\propto a$; we leave the analysis of the more general case where $P_z$ is non-negligible to future work.

It is straightforward to extend Eq.~\eqref{c:simple} to allow for smooth spatial and temporal dynamics of the order parameter. Taking into account the center of mass coordinate $\bm R$ of the excitons we obtain 
\begin{multline}
\label{c:fluct}
\mathrm i \hbar \frac{\partial a(\bm R, t)}{\partial t} = - \frac{\hbar^2}{2M}\Delta_{\bm R} a(\bm R, t)  + (E_g - E_B) a(\bm R, t)\\ + g |a(\bm R, t)|^2 a(\bm R, t) - d_{\rm exc} E^{\sigma^+}(\bm R, t),
\end{multline}
where $M=m_e + m_h$ is the exciton mass describing the center-of-mass motion and $\bm E(\bm R, t)$ is the electric field vector 
\begin{equation}
    \label{field}
    \bm E(\bm R, t) = \bm E_{\bm Q,\Omega}e^{\mathrm i \bm Q \bm R - \mathrm i \Omega t} + \bm E_{-\bm Q,-\bm \Omega} e^{- \mathrm i \bm Q \bm R + \mathrm i \Omega t},
\end{equation}
where $\bm E_{-\bm Q,-\bm \Omega} = \bm E_{\bm Q,\bm \Omega}^*$,
%\begin{subequations}
\begin{align}
        \label{circular}
%    &E^{\sigma^\pm}(\bm R, t)= \frac{E_x(\bm R, t) \mp \mathrm i E_y(\bm R,t)}{\sqrt{2}}, \\
    &E^{\sigma^\pm}_{\bm Q,\Omega} = \frac{E_{x;\bm Q,\Omega} \mp \mathrm i E_{y;\bm Q,\Omega}}{\sqrt{2}},
\end{align}
%\end{subequations} 
are the corresponding circularly polarized components.

Presenting $a(\bm R, t)$ in the form
\begin{equation}
\label{uv:def}
a(\bm R, t) = a e^{\mathrm i \chi} + f e^{\mathrm i \bm Q \bm R - \mathrm i \Omega t}+ w^* e^{2\mathrm i \chi -\mathrm i \bm Q \bm R + \mathrm i \Omega t},
\end{equation}
where $a$ satisfies Eq.~\eqref{c}, $\bm Q$ is the wavevector of the excitation, and $\Omega$ is its frequency, 
we obtain the set of equations for the amplitudes $f$ and $w$
\begin{align}
\hbar\Omega f = \frac{\hbar^2 Q^2}{2M}f + (E_B - E_g) (f+ w) -d_{\rm exc} E^{\sigma^+}_{\bm Q,\Omega},\\
-\hbar\Omega w = \frac{\hbar^2 Q^2}{2M}w + (E_B -  E_g) ( f+w) \nonumber\\
-d_{\rm exc}^* e^{2\mathrm i \chi} E^{\sigma^-}_{\bm Q,\Omega},
\end{align}

To determine the spectrum of elementary excitations in the absence of light-matter interaction we set $\bm E(\bm R, t)=0$ and arrive at the standard dispersion of the condensate excitations in the form 
\begin{equation}
\label{excitations}
\varepsilon(\bm Q) = \sqrt{\left(\frac{\hbar^2 Q^2}{2M}\right)^2 + \frac{\hbar^2 Q^2}{M}(E_B - E_g)}.
\end{equation}
In particular, for $Q\to 0$ the spectrum is sound-like 
\begin{equation}
    \label{sound}
    \varepsilon(Q) =\hbar c_s Q, \quad c_s = \sqrt{\frac{E_B -E_g}{M}},
\end{equation}
where $c_s$ is the effective Bogoliubov velocity (speed of sound in the condensate); the difference $E_B-E_g>0$ plays the role of the chemical potential of the condensate. We recall that in two dimensional systems such form of excitation spectrum prevents true condensation at finite $T$ (see Eqs.~\eqref{linear:bogolon:divergence} below). In the next section, we show that the allowance for the coupling of excitons with the vacuum electromagnetic field drastically changes the situation and allows for the true Bose-Einstein condensation. 

We now calculate the response of the exciton insulator with respect to the external electromagnetic field. Let us select the in-plane coordinate system in such a way that $x$-axis is aligned with the static polarization $\bm P$ which is equivalent to setting $\chi=0$. Introducing two response functions ($E_B \geqslant E_g$)
\begin{subequations}
    \label{bare:response}
    \begin{align}
        \pi_1(\Omega, \bm Q) = d_{\rm exc}^2\frac{E_B - E_g + \frac{\hbar^2 Q^2}{2M} + \hbar\Omega}{\varepsilon^2(\bm Q)-(\hbar\Omega)^2}, \label{pi1:condensate} \\
        \pi_2(\Omega, \bm Q) =  d_{\rm exc}^2 \frac{E_B - E_g }{ \varepsilon^2(\bm Q) - (\hbar\Omega)^2},
    \end{align}
\end{subequations}
we have 
\begin{subequations}
    \label{response:circ}
    \begin{align}
        P^{\sigma^+}_{\bm Q,\Omega}=\pi_1(\Omega,\bm Q) E^{\sigma^+}_{\bm Q,\Omega} - \pi_2(\Omega,\bm Q) E^{\sigma^-}_{\bm Q,\Omega},\\
    P^{\sigma^-}_{\bm Q,\Omega}=\pi_1(-\Omega,\bm Q) E^{\sigma^-}_{\bm Q,\Omega} -  \pi_2(\Omega,\bm Q)E^{\sigma^+}_{\bm Q,\Omega},
    \end{align}
\end{subequations}
where we took into account that oscillatory components of polarization are related to the coefficients $f$ and $w$ in Eq.~\eqref{uv:def} as $P^{\sigma^+}_{\bm Q,\Omega}= d_{\rm exc} f$, and $P^{\sigma^-}_{\bm Q,\Omega} = d_{\rm exc} w$. For completeness we note that in the absence of condensate $\pi_2(\Omega, \bm Q)\equiv 0$ while $\pi_1(\Omega, \bm Q)$ reduces to
\begin{equation}
    \label{pi1:normal}
    \pi_1(\Omega, \bm Q) = \frac{d_{\rm exc}^2}{E_g - E_B +  \frac{\hbar^2 Q^2}{2M} - \hbar\Omega}.
\end{equation}

Interestingly, the optical response of the system described by Eqs.~\eqref{response:circ} is anisotropic in the presence of the exciton condensate. Indeed, the right hand circularly polarized component of oscillatory polarization can be induced not only by the component of the incident electromagnetic fields with the same helicity, but also by the left hand circularly polarized fields (and vice versa). This polarization conversion effect is described by the response function $\pi_2(\Omega,\bm \Omega)$ which vanishes in the absence of the condensate. Physically, $\pi_2(\Omega,\bm \Omega)$ emerges due to the optical anisotropy of the system in the presence of the order parameter $a\ne 0$. This condensate-induced anisotropy of the system is also clearly seen in the linearly polarized basis; it is particularly strong in the small $\bm Q$, $\Omega$ regime where
\begin{equation}
    \label{small:OQ}
    \hbar\Omega,~\varepsilon(\bm Q),~\frac{\hbar^2Q^2}{2M} \ll E_B - E_g.
\end{equation}
Equation~\eqref{small:OQ} corresponds to the range of wavevectors and frequencies where the condensate effects are dominant. Here $\pi_1(\Omega, \bm Q) \approx \pi_2(\Omega, \bm Q)$ and, hence,
\begin{subequations}
    \label{response:lin:strong}
    \begin{align}
     P_{x;\bm Q,\Omega}& \approx -\mathrm i d_{\rm exc}^2 E_{y;\bm Q,\Omega}\frac{\hbar\Omega}{\varepsilon^2(\bm Q) - (\hbar\Omega)^2},\\
     P_{y;\bm Q,\Omega}&=2\pi_1(\Omega,\bm Q) E_{y;\bm Q,\Omega} \nonumber \\
     &\approx  2d_{\rm exc}^2  E_{y;\bm Q,\Omega} \frac{E_B - E_g }{ \varepsilon^2(\bm Q) - (\hbar\Omega)^2} .
    \end{align}
\end{subequations}
We recall that we use the frame of axes where the static polarization of the condensate $\bm P \parallel x$. Here the response along the $x$-axis is strongly suppressed, while the response along the $y$-axis is prominent. At fixed $E_y$ we have $|P_x/P_y| \sim \hbar\Omega/(E_B-E_g) \ll 1$. Electric fields polarized along the $x$-axis tend to change the amplitude of the order parameter, and such changes are energetically costly, while the field polarized along the $y$-axis tends to rotate the order parameter while preserving its magnitude; this rotation corresponds to variation of the overall condensate phase.

\section{Phase fluctuations and true condensation}\label{phase}

The long-wavelength low-frequency Goldstone mode associated with the phase fluctuations of the order parameter (Eq.~\eqref{excitations}) plays a crucial role in the response and stability of condensates. In particular, thermal fluctuations of the order parameter phase are known to destroy the long-range order in two-dimensional systems with broken continuous symmetries,\cite{PhysRevLett.17.1133,PhysRev.158.383} resulting in Berezinskii-Kosterlitz-Thouless physics for small but non-zero $T$.\cite{1971JETP...32..493B,1972JETP...34..610B,1973JPhC....6.1181K} Here we study phase fluctuations in the exciton insulator phase and demonstrate that for optically active excitons the coupling with the electromagnetic field suppresses the efficacy of thermal fluctuations and stabilizes a true Bose-Einstein condensate for $T > 0$.

We start with the analysis of the spectrum of elementary excitation of excitonic insulator by taking into account light-matter interaction. To this end, we complement the material relation in the form of Eq.~\eqref{response:lin:strong} by the solution of Maxwell's equations that relates the electric field with the polarization produced by the excitations of the condensate:
\begin{subequations}
    \begin{equation}
\label{E:Greens:2D}
E_{\alpha,\Omega} = \sum_{\beta} D^E_{\alpha\beta}(\Omega,\bm Q) P_{\beta;\bm Q,\Omega},  
\end{equation}
where the electromagnetic Greens function takes the form~\cite{ivchenko05a}
\begin{equation}
    \label{greens:EE}
    D^E_{\alpha\beta}(\Omega,\bm Q) =  \frac{2\pi  (\Omega/c)^2 }{\sqrt{Q^2-(\Omega/c)^2- \mathrm i 0}} \left(\delta_{\alpha\beta} - \frac{Q_\alpha Q_\beta}{(\Omega/c)^2}\right),
\end{equation}
\end{subequations}
Here, $c$ being the speed of light in the medium surrounding the TMD heterostructure. We once again use the coordinate frame where the static excitonic polarization is $\bm P \parallel x$. We study the long-wavelength excitations and keep the leading order in $\hbar\Omega \ll E_B - E_g$ terms in the material relation~\eqref{response:lin:strong}: in this limit only $P_{y,\bm Q,\Omega}$ is non-zero and the self-consistency equation reads
\begin{equation}
    \label{spec:self:gen}
    2\pi_1(\Omega,\bm Q) D_{yy}^E(\Omega,\bm Q) = 1.
\end{equation}
%\commentAtac{This follows the fact that $P_y \gg P_x$, but do we need to consider the term from $D_{yx}$ or is this a small correction?} \commentMisha{I have checked, it is a small correction which does not change the results much. It is because our system (one valley) supports only one polarization (fixed linear at $\Omega, \bm Q=0$ which transforms to a circular for high $\Omega,\bm Q$).}
This equation determines the dispersion of the elementary excitations of the excitonic insulator $\mathcal E(\bm Q)$ when light-matter interaction is fully taken into account. In the absence of condensate, our results (see Appendix~\ref{sec:exc:LT:standard}) are in complete agreement with the previous literature \cite{ivchenko05a,glazov2014exciton}.

\begin{figure}[t!]
    \centering
    \includegraphics[width=\linewidth]{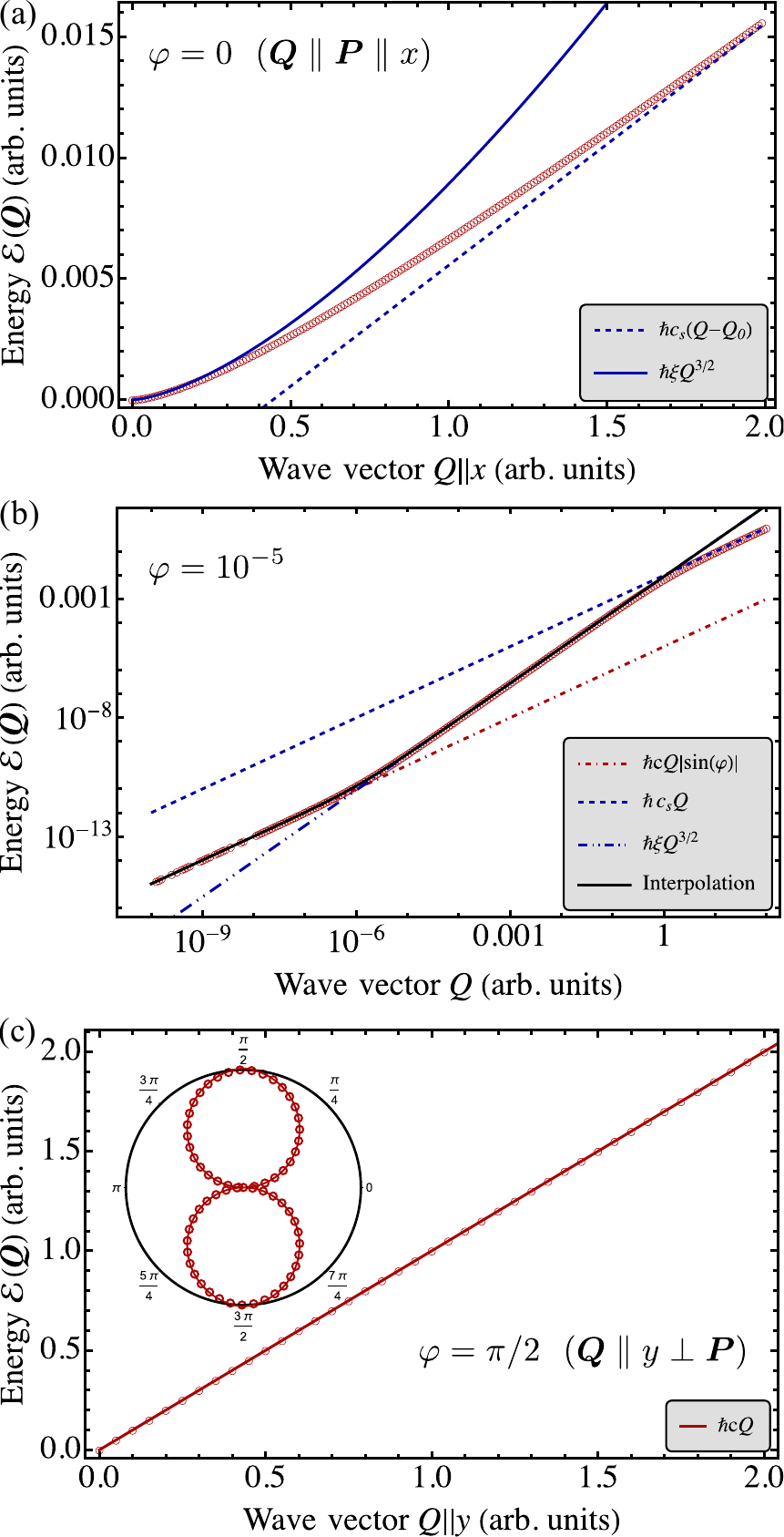}
    \caption{Dispersion of elementary excitations of excitonic insulator with allowance for the light matter coupling. Open symbols show the dispersion calculated numerically, solid lines show analytical approximations; $x$-axis corresponds to the polarization of the condensate $\bm P \parallel x$, see text for details. (a) $\bm Q\parallel x$ ($\varphi=0$). Dashed curve with $\addMisha{c_s}Q$ asymptotics is shifted to allow for the $Q$-independent overall shift. (b) Small, but non-zero $\varphi=10^{-5}$. Dot-dashed and dashed curves show $Q$-linear asymptotics~\eqref{light:like} and \eqref{tr2}, blue dot-dot-dashed line shows $Q^{3/2}$ law, Eq~\eqref{tr1}, and black solid line shows the interpolation~\eqref{disper:interp}.  (c) $\bm Q\parallel y$ ($\varphi=\pi/2$). Inset shows the angular dependence of the energy with the curve calculated after Eq.~\eqref{light:like}. }
    \label{fig:disper}
\end{figure}

At $Q\to 0$ the dispersion is
\begin{subequations}
    \label{dispersion:cases}
    \begin{equation}
    \label{light:like}
        \Omega \equiv \mathcal E(\bm Q)/\hbar \approx c |Q_y|, %- c\frac{|Q_xQ_y|}{8\pi d_{\rm exc}^2}
        \quad |Q_y|> \frac{c_s}{c} Q,
    \end{equation}
where $c_s$ is defined in Eq.~\eqref{sound}. This expression is valid for practically all directions of $\bm Q$ apart from a small range of angles $\varphi=\angle \bm Q, \bm x$, $\varphi<c_s/c$, where, for fixed $Q$, the $y$-component of the wavevector becomes particularly small. For $\varphi=0$ (i.e. $\bm Q\parallel x$) we find
\begin{align}
    \Omega \equiv \mathcal E(\bm Q)/\hbar= \xi Q^{3/2}, \quad Q \ll \frac{c_s^2}{\xi^2},     \label{tr1}\\
    \Omega \equiv \mathcal E(\bm Q)/\hbar= c_sQ_x, \quad Q\gg \frac{c_s^2}{\xi^2} \label{tr2},
\end{align}
\end{subequations}
where $\xi = \hbar c_s c/[4\pi d_{\rm exc}^2(E_B - E_g)]^{1/2}$.
The dispersion of elementary excitations is shown in Fig.~\ref{fig:disper} with the focus on three key cases: (a) $\varphi=0$ ($\bm Q\parallel x$), (b) the very small but non-zero case $\varphi$ where $c|Q_y| \ll c_s Q$, and (c) the case of $\varphi=\pi/2$ ($\bm Q \parallel y$) are presented in the respective panels of Fig.~\ref{fig:disper}.

%%%%%%%%%%%%%%%%%%%%%%%%AI%%%%%%%%%%%%%%%%%%%%

Naturally, if $|Q_y|$ is large enough the spectrum is linear and close to $c Q_y$ asymptotics in Eq.~\eqref{light:like}. This is a  consequence of the fact that for sufficiently large $|Q_y|$ the excitations have significant longitudinal component of the dipole moment whose coupling with electromagnetic near field pushes their energy to light cone (see Fig.~\ref{fig:disper}(c)). This situation is similar to the case the two-dimensional plasmon in the presence of retardation.\cite{Stern67a,Chaplik:1985aa} We note that linear dispersion is generic for two-dimensional longitudinal excitons in quantum well structures and in atomically thin semiconductors; for excitons with a finite energy $E_x({\bm Q} = 0) = E_g - E_B$ however, linear dispersion is only observed for states outside the light cone (Appendix~\ref{sec:exc:LT:standard}).\cite{maialle93,goupalov98,Yu:2014fk-1,glazov2014exciton}
Interestingly, even if $|Q_y|$ is small, the dispersion at $Q\to 0$ is still dominated by the light matter interaction, as illustrated in Fig.~\ref{fig:disper}(a,b). For instance, at $\varphi=0$ the dispersion starts as $Q^{3/2}$ and turns over to the $c_sQ$ law typical for the Bogoliubov excitations only for  wavevectors which exceed $Q_* \approx c_s^2/\xi^2$ (see Eqs.~\eqref{tr1} and \eqref{tr2}). For $|\varphi|\lesssim {c_s}/c$ the dispersion, as illustrated in Fig.~\ref{fig:disper}(b) starts as $c|\sin{\varphi}|Q$ then turns to $Q^{3/2}$ asymptotics~\eqref{tr1}, and with further increase in $Q$ approaches the linear law \eqref{tr2} dominated by the interactions in the condensate. In the relevant regime of small and intermediate $ Q$ the interpolation function
\begin{equation}
    \label{disper:interp}
    \mathcal E^2(\bm Q) = \hbar^2 c^2 Q^2\sin^2{\varphi} + \hbar^2\xi^2 Q^{3}, \quad \varphi\lesssim {c_s}/c,
\end{equation}
rather accurately describes the results of numerical calculations, as shown by the black curve in Fig.~\ref{fig:disper}(b). This nontrivial dispersion of Bogoliubov excitations is a result of the coupling of excitons to the electromagnetic field. The dispersion derived in Eqs.~\eqref{spec:self:gen} and shown in Fig.~\ref{fig:disper} deviates from the generic Bogoliubov dispersion given by Eq.~\eqref{sound} and indicates that the condensate fluctuations are qualitatively different.

The strong modification of the collective mode dispersion is somewhat similar to the Anderson-Higgs effect in superconductors. \cite{PhysRev.112.1900,PhysRevB.26.4883,annurev:/content/journals/10.1146/annurev-conmatphys-031214-014350,Anderson:2015aa,Altland_Simons_2010} We can understand the origin of this modification by considering the limit $|E_B - E_g| < kT$ and neglecting the exciton-exciton interactions and light-matter coupling: the elementary excitations in this limit would be characterized by a parabolic dispersion, Fig.~\ref{fig:excit:simple}(a). Upon taking into account coupling of excitons to the electromagnetic field, the longitudinal exciton dispersion would change from parabolic to linear (Eq.~\eqref{E:0:L}), indicating that condensation is possible. When $T$ is such that a condensate mean-field description is justified, the gapless collective Bogoliubov mode  will have a $Q^{1/2}$ dispersion (cf. Ref.~\onlinecite{PhysRevLett.131.236004}). Although this consideration indicates the possibility of a true long-range order, the aforementioned $Q^{1/2}$ dispersion is not physical, since it lies above the light cone and violates causality. The proper analysis we have carried out shows that the excitations in the low $Q$ excitations become light-like and approach the light cone at $Q\to 0$. Correspondingly, these low energy excitations contain a vanishing fraction of exciton character in the $Q\to 0$ limit, which in turn ensures suppression of thermal depletion of the condensate.

%In a simplified picture one can understand this result as follows. Consider an onset of condensation where $E_B$ reaches the $E_g$. Hence, we start with non-interacting excitons characterized by the parabolic center of mass dispersion, Fig.~\ref{fig:excit:simple}(a). Let us first take into account the light-matter interaction. As shown in Appendix~\ref{sec:exc:LT:standard} the longitudinal exciton dispersion will change from parabolic to linear, Eq.~\eqref{E:0:L}. The condensation of excitons with linear dispersion should, following Bogolyubov's approach result in the collective mode with the $Q^{1/2}$ dispersion. Importantly, such dispersion lies above the light cone and, hence, violates causality. The proper analysis (as we have done above) shows that the excitations in the system become very light-like and approaches the light cone at $Q\to 0$. This argument also shows that the elementary excitations are light-like and contain just a small fraction of excitons in the $Q\to 0$ limit. As we see below, it results in suppression of thermal depletion of the condensate.

To verify that the thermal fluctuations do not destroy the condesate, we construct the Greens function of the center-of-mass motion of the excitons.  In the presence of the light-matter coupling the exciton susceptibility with respect to the electromagnetic field is renormalized compared to $\pi_1(\Omega, \bm Q)$ due to the self-consistent interaction
\begin{equation}
\label{Greens:EI}
\mathcal G(\Omega, \bm Q) =- \frac{1}{d_{\rm exc}^{2}} \frac{\pi_1(\Omega, \bm Q)}{1-  2\pi_1(\Omega,\bm Q) D_{yy}^E(\Omega,\bm Q)}.
\end{equation}
Naturally, the poles of $\mathcal G$ correspond to the solutions of Eq.~\eqref{spec:self:gen}.
The number of thermally excited excitons  is given by\cite{ll9_eng}
\begin{equation}
\label{n:exc:EI:L:branch}
N_{\rm exc} = \sum_{\bm Q} \int \frac{d\Omega}{2\pi \mathrm i} \left[\mathcal G(\Omega, \bm Q)  - \mathcal G^*(\Omega, \bm Q)\right] n(|\hbar\Omega|),
\end{equation}
where 
\begin{equation}
n(\hbar\Omega) = \frac{1}{\exp{(\hbar\Omega/k_B T)} - 1} \approx \frac{k_B T}{\hbar \Omega},
\end{equation}
is the quasiparticle distribution function at $T$.  The condensation in the strict sense of the term is possible if the integral converges at $Q\to 0$, i.e., if excited states host finite number of particles. To proceed, we express the Greens function in the vicinity of its poles as
as
\begin{equation}
    \label{G:lm:0}
    \mathcal G(\Omega, \bm Q) = \frac{\mathcal A(\bm Q)}{\Omega^2 - \mathcal E^2(\bm Q) - \mathrm i 0},
\end{equation}
with $\mathcal A(\bm Q)$ denoting the quasi-particle weight. With this approximation, we find
\begin{equation}
    \label{N:exc:gen}
    N_{\rm exc} = \sum_{\bm Q} \frac{\mathcal A(\bm Q)}{\mathcal E(\bm Q)}n[\mathcal E(\bm Q)] \approx k_B T\sum_{\bm Q} \frac{\mathcal A(\bm Q)}{\mathcal E^2(\bm Q)},
\end{equation}
where in the last equality, the summation over $\bm Q$ should be cut-off at $\mathcal E(Q)\sim k_B T$. To verify that  $N_{\rm exc}$ does not diverge, we derive an expression for $1/ \mathcal A(\bm Q)$ in the limit of small $Q$ and $\varphi$
\[
\frac{1}{\mathcal A(\bm Q)} = 1+ 2 d_{\rm exc}^2 \left.\frac{\partial}{\partial (\hbar\Omega)^2} D_{yy}^E(\Omega,\bm Q)\right|_{\Omega = \mathcal E(\bm Q)}.
\]
The quasi-particle weight $\mathcal A(\bm Q)$  quantifies the exciton character of the elementary excitations. In the relevant range of small $Q$ and $\varphi$ the factor $\mathcal A^{-1}=4\pi d_{\rm exc}^2/Q$ yields

\begin{equation}
    \label{G:lm}
    \mathcal G(\Omega, \bm Q) =\frac{(\hbar c)^2}{4\pi d_{\rm exc}^2} \frac{Q}{(\hbar\Omega)^2 - \mathcal E^2(\bm Q) - \mathrm i 0}.%,
\end{equation}
We emphasize that the factor $Q$ appearing in the numerator of $\mathcal G(\Omega, \bm Q)$ indicates the vanishing quasi-particle weight  $\mathcal A(Q)\to 0$ of the collective excitations as $Q\to 0$. Using the small-$Q$ and small-$\varphi$ asymptotics, we obtain
\begin{equation}
    \label{N:exc:conv}
    N_{\rm exc} = \frac{c k_B T}{\sqrt{3}(2\pi)^2 d_{\rm exc}^2} \times  \left(\frac{ k_B T}{\hbar\xi^4} \right)^{1/3}. 
\end{equation}
 Remarkably, there is a finite number of thermally excited excitons at finite $T$, demonstrating that light-matter coupling ensures Bose-Einstein condensation of ground state interlayer excitons. 
 %By contrast to the three-dimensional systems where $N_{\rm exc}^{3D} \propto T^2$ and two-dimensional systems in the absence of light-matter coupling where $N_{\rm exc}$ diverges, here $N_{\rm exc}\propto T^{4/3}$.

The aforementioned results stands in contrast  to what one finds in the absence of the light-matter interaction ($d_{\rm exc} \equiv 0$): It follows from Eqs.~\eqref{pi1:condensate} and \eqref{Greens:EI} that
\begin{subequations}
    \label{linear:bogolon:divergence}
\begin{equation}
\mathcal G(\Omega, \bm Q) \approx \frac{E_B - E_g}{(\hbar\Omega)^2 - c_s^2Q^2 - \mathrm i 0},
\end{equation}
which for two-dimensional excitons gives
\begin{equation}
N_{\rm exc} \propto \sum_{\bm Q} \frac{k_B T}{Q} \frac{1}{Q} \to \mbox{diverges,}
\end{equation}
\end{subequations}
indicating that thermal fluctuations deplete the condensate.\cite{ll9_eng}

\section{Experimental signatures}

The search for the smoking gun evidence of the excitonic insulator formation is one of the outstanding challenges in the field. As a necessary but not sufficient first step, one could probe the presence of spontaneously generated ground-state excitons through detection of the $1s-2p$ dipole-active intra-exciton transition.\cite{Poellmann:2015aa} We envisage a simple experiment where the THz absorption spectrum is measured as a function bias voltage which varies $E_g-E_B$: the onset of the THz absorption demonstrates the spontaneous formation of excitons in the system.

\begin{figure*}[tb]
    \centering
    \includegraphics[width=0.75\linewidth]{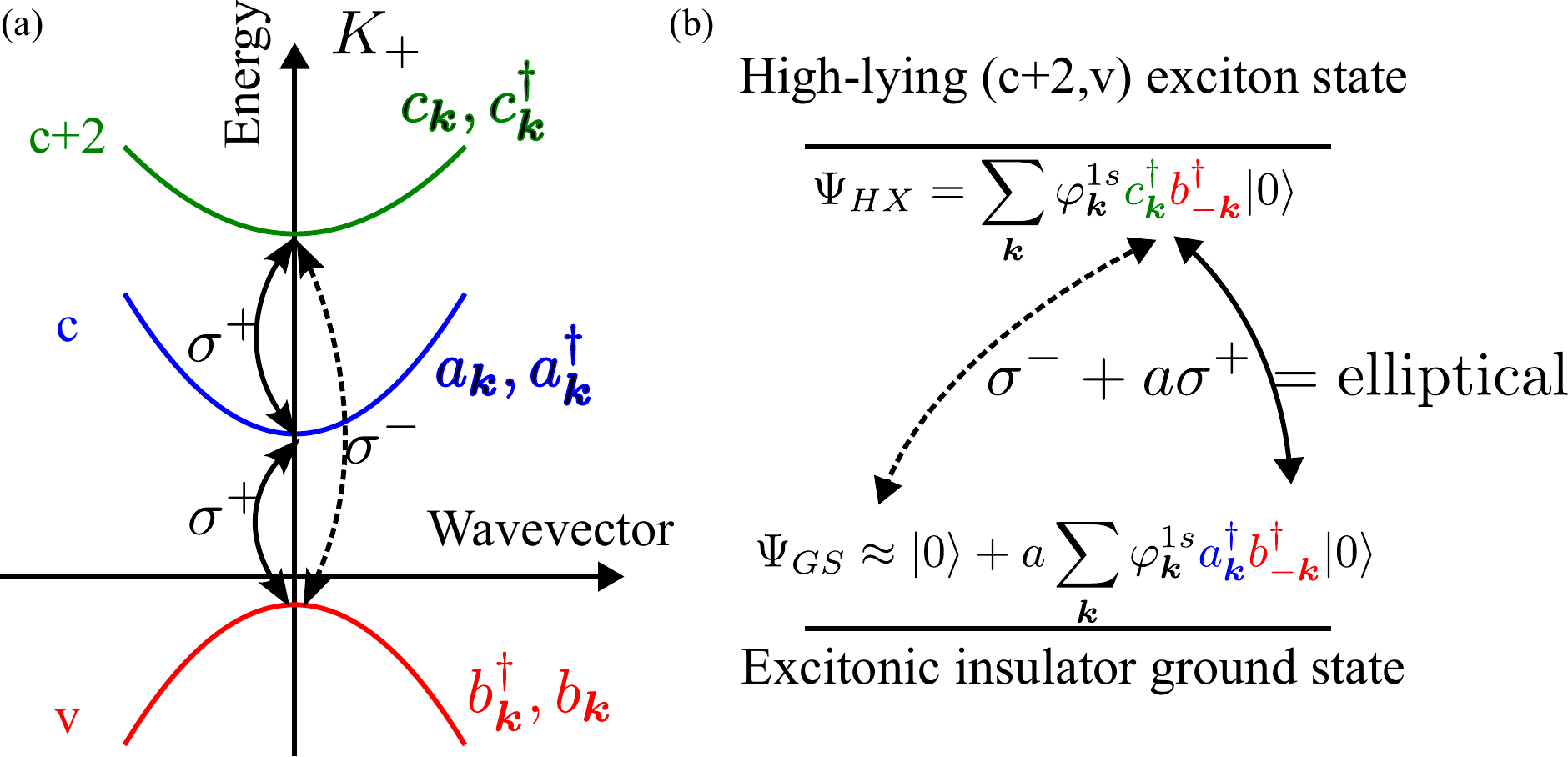}
    \caption{Detection of the exciton insulator state with the ferroelectric order. (a) Schematics of the TMD heterobilayer bandstructure in the $K_+$ valley that shows in addition to the nearest valence and conduction band the remote $c+2$ band of the opposite to the $c$-band chirality. The selection rules for the interband transitions are shown by arrows. (b) Interference of transitions for excitation or recombination of the high-lying exciton associated with the remote conduction band. In the presence of excitonic condensate, $a\ne 0$, the selection rules for such transition involve elliptical polarization.}
    \label{fig:highlying}
\end{figure*}

Generally, the formation of the ordered state can be detected using optical spectroscopy since ferroelectric order should change the selection rules for transitions from purely circular to elliptical.  Figure~\ref{fig:highlying} details the generic TMD band structure where $c+2$ remote conduction band  has an opposite $C_3$ angular momentum compared to the conduction band~\cite{2053-1583-2-2-022001,PhysRevB.95.035311}. The optical transitions from the valence band $v$ to the $c+2$ conduction band of the opposite monolayer in the $K_+$ valley are $\sigma^-$ circular polarized while the transitions from the bottom conduction band $c$ to $c+2$ in the same monolayer  $\sigma^+$ circularly polarized (Fig.~\ref{fig:highlying}(a)). The so-called ``high-lying'' excitonic transitions associated with this remote band have already been observed and studied both for mono- and bilayer structures\cite{Lin:2021uu,Lin:2022aa,Lin:2024aa}.

The ground state in the presence of and excitonic insulator and ferroelectric order is a mixture of the “vacuum” state $|0\rangle$ and the  exciton Fock states $|1\rangle, |2\rangle$, etc. (Eq.~\eqref{coherent}). If we focus on the 1s exciton transition between the ground state in Eq.~\eqref{coherent} and the $c+2$ band as shown in Fig.~\ref{fig:highlying}(b), we notice that there are two pathways for this transition: from $|0\rangle$ by $\sigma^-$ photon or from $|1\rangle$ by $\sigma^+$ photon. Consequently, this transition will generally be elliptically polarized and the polarization axis will be linked to the phase of the condensate’s order parameter $a$. The magnitude of linear polarization is proportional to $|a|$. The oscillator strength of the transition $v\leftrightarrow c+2$ (active in the $\sigma^-$ polarization) can be relatively weak compared to that of $c\leftrightarrow c+2$, since it involves the interlayer transfer of the hole. However, in the regime where condensate order parameter $|a|$ is small as well, the degree of elliptical polarization of the optical transition can be quite sizable.

The emergence of ferroelectric order in interlayer excitons may also be detected by intra-layer exciton spectroscopy. It has been predicted and experimentally demonstrated that the presence of interlayer excitons leads to the formation of attractive and repulsive polaron branches in optically active intralayer excitonic resonances.\cite{amelio_polaron_2023} The attractive polaron (AP) in this case, can be considered as a collective excitation of biexcitons -- bound state of an intra- and interlayer exciton~\cite{amelio_polaron_2023}.  By measuring the polarization dependence of the AP branch in each layer, it is also possible to determine the valley degree of freedom of both the electron and the hole forming the interlayer exciton; for example, if the interlayer excitons are bound $K_+$-valley electron-hole pairs, then the AP resonances of both layers would be right-hand circularly polarized.

Upon condensation and formation of the ferroelectric order, we could expect the polarization properties of AP resonances to be modified. In the limit where the biexciton radius is smaller than the average separation between ground-state interlayer excitons, the biexcitons would have a finite electric dipole moment along the same axis as the condensate, which should lead to linearly polarized AP resonance. The effect is related to previous reports of linearly polarized excitonic resonances observed upon formation of nematic order of electrons~\cite{Xu2020}.  By measuring the linear polarization axis of AP across the sample, it should be possible to determine the correlation length of the ferroelectric order.

\section{Outlook}

Dipolar 2D interlayer excitons in van der Waals heterostructures have emerged as a new platform for investigation of strongly correlated bosons in the solid state. While experiments have already demonstrated  perfect drag and Wigner crystal states of finite-energy excitons, outstanding theoretical proposals range from exciton mediated superconductivity~\cite{Zerba2023,vonmilczewski2023} to bosonic fractional Chern insulators.\cite{Hafezi-PhysRevLett.133.136403,Raul2025excitonfractionalcherninsulators}  Even though in the short run the experiments are likely to be carried out in small systems where the distinction between the power-law decay of the correlations and the true long-range order of the parent excitonic insulator state would not be discernible, our findings are important not only for future experiments but also for our understanding of the possibilities in the exciting platform of 2D materials.

We also highlight a connection to 2D XXZ magnets, where it is well known that a ferromagnetic phase transition that breaks the U(1) symmetry is possible due to flip-flop magnetic dipole interactions. The precise connection between the two problems, as well as the role of long-range dipolar interactions between ground-state excitons will be the subject of future work.

Even though vacuum electromagnetic field fluctuations lead to observable effects such as the Lamb shift or the Casimir effect, an interesting open question in condensed matter physics is whether the functionality of a quantum material can be modified through coupling to cavity enhanced vacuum fields~\cite{Ashida-PhysRevX.10.041027}. While we do not assume a cavity structure, we show here that the coupling of elementary excitations to the vacuum field leads to a qualitative change in the ground state of a 2D material by inducing a paraelectric-to-ferroelectric phase transition. Naturally, enhancement of vacuum fluctuations at vanishing energies would lead to a higher $T_c$ and a more robust ferroelectric state, similar to the proposals of Higgs mode stabilization in superconductors\cite{PhysRevB.104.L140503}.

\section{Conclusion}

We showed that true long-range ferroelectric order and Bose-Einstein condensation of interlayer excitons can occur in two-dimensional heterobilayer semiconductors, despite the restrictions imposed by the Mermin-Wagner theorem. The key requirements include reducing the interlayer band gap below the exciton binding energy via an applied electric field, applying a finite magnetic field to lift the valley degeneracy, and having a band structure that supports long-range electron-hole exchange interaction. Under these conditions, excitons spontaneously form in the ground state and acquire a photon-like linear dispersion at small momenta due to coupling with the electromagnetic field, which suppresses phase fluctuations and stabilizes the condensate at finite temperatures.

The excitonic insulator phase is characterized by spontaneous breaking of U(1) symmetry, leading to a ferroelectric state with in-plane electric polarization. The coexistence of superfluidity and ferroelectricity in this system represents a qualitatively new electronic ground state enabled by vacuum fluctuations. The findings establish dipolar 2D interlayer excitons as a promising platform for exploring strongly correlated bosonic phases and suggest that cavity-enhanced vacuum fields could further stabilize the ferroelectric order, opening new possibilities for controlling quantum material functionality through light-matter coupling.

\subsection*{Acknowledgements}
The authors acknowledge insightful discussions with Allan MacDonald, Peter Littlewood, Andrew Millis, Ivan Iorsh and Eugene Demler. They also thank Anna Seiler for sharing Figure 1. 

%Here we could make connections to 2D XXZ magnets with long range dipole-dipole interactions and discuss possibilities for experimental observation.

\appendix

\section{Exciton instability at finite temperature}\label{sec:gap:T}

We consider a simplified version of a pairing for the short-range interaction, $V_{\bm k- \bm k'} \equiv V_0$ and symmetric dispersions of electrons and holes $E_e(\bm k) = E_h(-\bm k)$. In this case $\Delta_{\bm k} = \Delta$. The gap equation with allowance for the finite $T$ can be written as [cf. Ref.~\onlinecite{ll9_eng}]
\begin{equation}
    \label{gap:np}
    \frac{1}{V_0} = \sum_{\bm k} \frac{1-2n_{\bm k}}{2\xi_{\bm k}},
\end{equation}
where $$\xi_{\bm k} = \mathcal E_{\bm k}/2 = \sqrt{E_e^2(\bm k) + \Delta^2},$$ and $$n_{\bm k} = \frac{1}{\exp{(\xi_{\bm k}/k_B T)}+1},$$
where $k_B T$ is the temperature expressed in the units of energy.
Removing high-energy divergencies in Eq.~\eqref{gap:np} in a standard way and expressing $V_0$ via the bound state binding energy $E_B$ we arrive at the gap equation in the following form:
\begin{multline}
    \label{gap:np:1}
    \ln{\left(E_B\frac{\sqrt{E_g^2+4\Delta^2}-E_g^2}{2\Delta^2}\right)} \\
    = 2\int_0^\infty \frac{dx}{\sqrt{(E_g/2+x)^2+\Delta^2}\exp{\left(\frac{\sqrt{(E_g/2+x)^2+\Delta^2}}{k_B T}\right)}+1}.
\end{multline}

\begin{figure}[hbt]
    \centering
    \includegraphics[width=\linewidth]{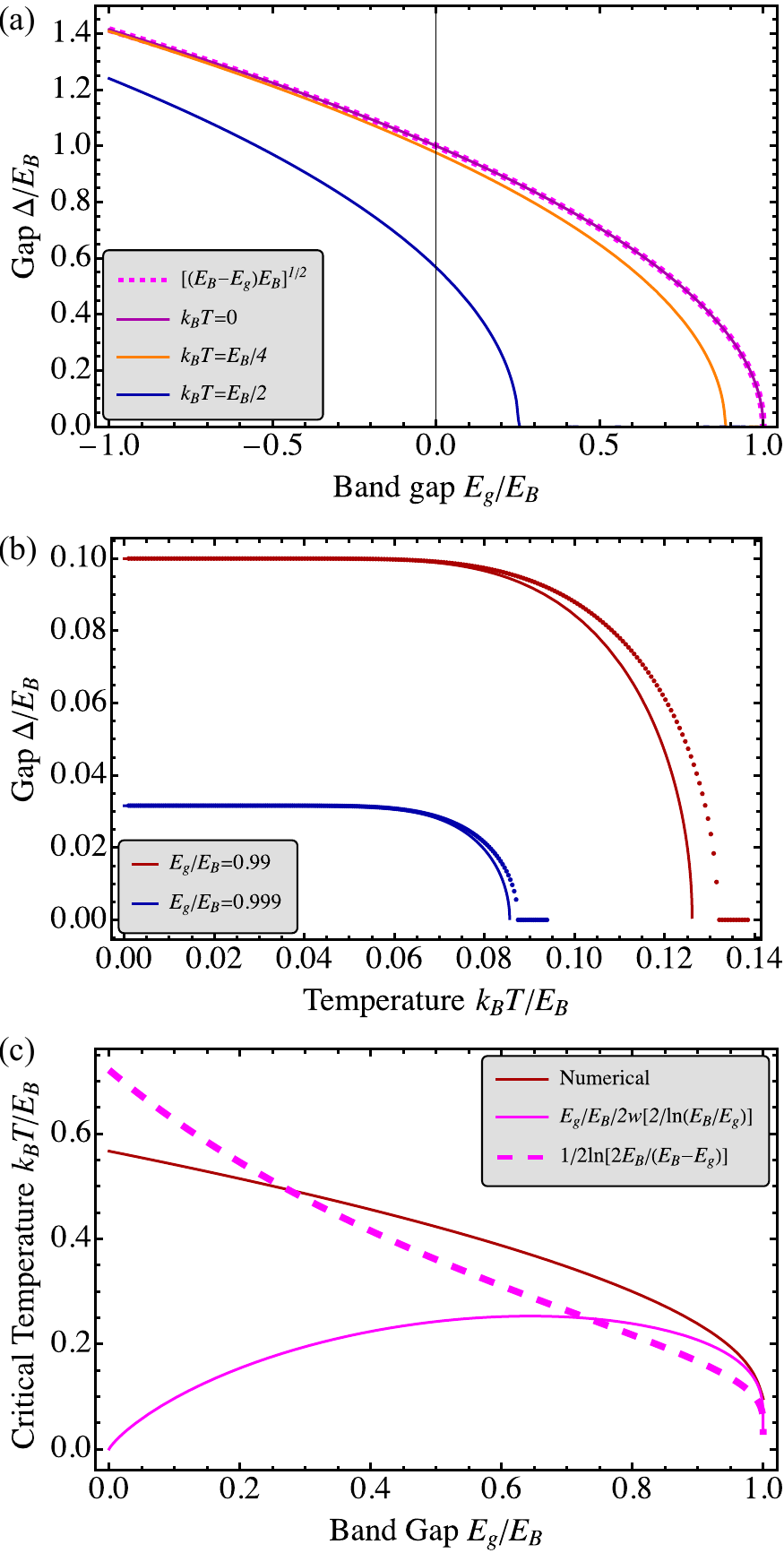}
    \caption{Pairing gap $\Delta$ as a function of the ratio $E_g/E_B$ (a) and temperature $k_B T$. (c) Critical temperature for the pairing $T_{\rm pair}$ as a function of the ratio $E_g/E_B$. See legends and text for details. }
    \label{fig:gap}
\end{figure}

At $T=0$ the right-hand side of Eq.~\eqref{gap:np:1} vanishes. For small $E_B-E_g \ll E_B$ the gap, $\Delta \ll E_B, E_g$ (hereafter in this section we select arbitrary phase of the gap to be zero for definiteness). Keeping the leading order terms in $\Delta/E_B$ we obtain
\begin{equation}
    \label{Delta:EBEg:simple}
    \Delta = \begin{cases}
        \sqrt{E_B(E_B - E_g)}, \quad E_B>E_g,\\
        0, \quad\mbox{otherwise}.
    \end{cases}
\end{equation}
This result is in agreement with Eq.~\eqref{c} of the main text. This dependence is illustrated in Fig.~\ref{fig:gap}(a): Solid red curve shows the gap dependence found numerically from Eq.~\eqref{gap:np:1} and dotted magneta curve shows Eq.~\eqref{Delta:EBEg:simple}.

In the relevant temperature range $k_B T \ll E_g, E_B$. It allows to evaluate integral in the right-hand side of Eq.~\eqref{gap:np:1} as
\begin{multline}
    \label{I3}
    2\int_0^\infty \frac{dx}{\sqrt{(E_g+x)^2+\Delta^2}\exp{\left(\frac{\sqrt{(E_g+x)^2+\Delta^2}}{k_B T}\right)}+1} \\
    \approx  -2{\rm Ei}(-E_g/k_B T) \approx \frac{2k_B T}{E_g} e^{-E_g/k_B T},
\end{multline}
where ${\rm Ei}(z) = -\int_{-z}^\infty e^{-t}/t dt$ is the exponential integral. Naturally, at finite $T$ the gap is reduced and the instability point shifts to smaller values of $E_g$ compared to the zero $T$-case of $E_B = E_g$. The gap as a function of $T$ is shown in Fig.~\ref{fig:gap}(b) where dots demonstrate full numerical solution of Eq.~\eqref{gap:np:1} and the solid lines represent the analytical result obtained using the second approximation in Eq.~\eqref{I3}
\begin{equation}
    \label{Delta:T:simple}
    \Delta = \sqrt{2\tilde E_B(T)(\tilde E_B(T) - E_g)},
\end{equation}
where 
\[
\tilde E_B(T) = E_B \exp{\left(\frac{4k_B T}{E_g} e^{-E_g/2k_B T} \right)}.
\]
This approximation works the better, the closer $E_g$ to $E_B$ is, i.e., the smaller the gap is.

Setting $\Delta=0$ we obtain from Eq.~\eqref{gap:np:1} the critical temperature of pairing, solid red line in Fig.~\ref{fig:gap}(c). Analytical approximations found from Eq.~\eqref{Delta:T:simple}: $E_B(T_{\rm pair})=E_g$ are shown by the solid magneta line (where the solution of this equation is expressed via the Lambert-$w$ function or product logarithm) and dotted magneta line where a crude logarithmic approximation
\begin{equation}
    \label{Tc:log}
    T_{\rm pair} = \frac{E_g}{2k_B} \frac{1}{\ln[2E_B/(E_B-E_g)]},
\end{equation}
has been employed. It is noteworthy that the critical temperature evaluated in this way corresponds to the critical temperature of the pairing. There are fluctuations of the phase of $\Delta$ (bosonic Goldstone mode) which affect the condensate and would destroy the long-range order in the system in the absence of coupling to the electromagnetic field (see the main text).

\section{Derivation of gap equation by unitary transform}\label{sec:gap:alternative}

Following Ref.~\onlinecite{GuseinovKeldysh} it is convenient to perform a transformation to the novel operators
\begin{subequations}
\label{Bog:1}
\begin{align}
a_{\bm k} \to u_{\bm k} a_{\bm k} + v_{\bm k} b_{-\bm k}^\dag, \quad & \quad a_{\bm k}^\dag \to u_{\bm k} a_{\bm k}^\dag + v_{\bm k}^* b_{-\bm k},\\
b_{{-}\bm k} \to u_{\bm k} b_{{-}\bm k} - v_{\bm k} a_{\bm k}^\dag,\quad & \quad b_{{-}\bm k}^\dag \to u_{\bm k} b_{{-}\bm k}^\dag - v_{\bm k}^* a_{\bm k},
\end{align}
where 
\[
\quad u_{\bm k}^2 + |v_{\bm k}|^2 =1,
\]
and it is convenient to parametrize $u_{\bm k}$ and $v_{\bm k}$ via the so far unknown function $\varphi(\bm k)$
\begin{multline}
\label{uv:varphi}
u_{\bm k} = \cos{\varphi(\bm k)} \approx 1- \frac{|\varphi(\bm k)|^2}{2},\\
v_{\bm k} = \mathrm i e^{\mathrm i \arg{\varphi(\bm k)}} \sin{\varphi(\bm k)} \approx \mathrm i \left( 1- \frac{|\varphi(\bm k)|^2}{6}\right)\varphi(\bm k).
\end{multline}
It is instructive to present
\begin{align}
\label{uv:prod:varphi}
u_{\bm k} v_{\bm k} \approx  \mathrm i \varphi(\bm k)\left[1-\frac{2}{3}|\varphi(\bm k)|^2 \right], \\ u_{\bm k}^2 - |v_{\bm k}|^2 \approx 1- 2|\varphi(\bm k)|^2
\end{align}
Note that the actual unitary transformation is provided by the operator
\[
S = \exp{\left\{\mathrm i\sum_{\bm k} \left[\varphi(\bm k) a^\dag_{\bm k} b^\dag_{-\bm k} - \varphi^*(\bm k) b_{-\bm k} a_{\bm k}  \right] \right\}}.
\]
The choice of real $u_{\bm k}$ is a matter of convenience. Function $\varphi(\bm k)$ plays a role of the order parameter, it describes the electron-hole correlations in the system's ground state.
\end{subequations}

The transformed Hamiltonian takes the form up to $\propto \varphi^3(\bm k)$ terms (operator-independent part is omitted)
\begin{multline}
\label{SHS}
\mathcal H = S^\dag (\mathcal H_2 + \mathcal V) S = \sum_{\bm k} \mathcal E_e(\bm k) a^\dag_{\bm k} a_{\bm k} + \mathcal E_h (\bm k) b^\dag_{\bm k} b_{\bm k} \\
+ \mathcal T + \mathcal M + \tilde{\mathcal V},
\end{multline}
with the renormalized electron and hole dispersions
\begin{subequations}
\label{ren:disper}
\begin{align}
\mathcal E_e(\bm k) =  E_e(\bm k) - |\varphi(\bm k)|^2 \left[E_e(\bm k) + E_h({-}\bm k)\right] \nonumber \\+ \left(u_{\bm k} v_{\bm k}^* \gamma_\alpha k_\alpha  - u_{\bm k} v_{\bm k} \gamma_\alpha^* k_\alpha\right) - \ldots,\\
\mathcal E_h(\bm k) = E_h(\bm k) - |\varphi(\bm k)|^2 \left[E_e({-}\bm k) + E_h(\bm k)\right] \nonumber \\ 
{-} 
\left(u_{\bm k} v_{\bm k}^* \gamma_\alpha k_\alpha - u_{\bm k} v_{\bm k} \gamma_\alpha^* k_\alpha\right) - \ldots,
\end{align}
\end{subequations}
renormalized interband coupling
\begin{multline}
\label{STS}
\mathcal T = \sum_{\bm k} \gamma_\alpha k_\alpha \left(1- |\varphi(\bm k)|^2 \right)a^\dag_{\bm k} b_{-\bm k}^\dag \\
+ \gamma_\alpha^* k_\alpha \varphi^2(\bm k)a^\dag_{\bm k} b_{-\bm k}^\dag + {\rm h.c.} \\
= \sum_{\bm k} \lambda_{\bm k} a^\dag_{\bm k} b_{-\bm k}^\dag  + \lambda_{\bm k}^* b_{-\bm k} a_{\bm k} ,
\end{multline}
renormalized interaction $\tilde{\mathcal V}$ containing four normally ordered creation and annihilation operators
%\begin{equation}
%\label{SVS}
%\tilde{\mathcal V} = \ldots,
%\end{equation}
and, finally, ``dangerous'' terms which contain two creation or two annihilation operators stemming from both the dispersion and interactions
\begin{equation}
\label{M}
\mathcal M = \sum_{\bm k} M_{\bm k} a^\dag_{\bm k} b^\dag_{-\bm k}+  M_{\bm k} b_{-\bm k} a_{\bm k},
\end{equation}
where
%\begin{widetext}
\begin{multline}
\label{Mk}
M_{\bm k} 
=  \underbrace{\mathrm i \varphi(\bm k)\left[1-\frac{2}{3}|\varphi(\bm k)|^2 \right]  \left[E_e(\bm k) + E_h({-}\bm k)\right]}_{\mbox{\tiny free particle dispersion} \propto u_{\bm k} v_{\bm k}} \\
-\mathrm i \sum_{\bm k'}\underbrace{ V_{\bm k - \bm k'} \left[1-2 |\varphi(\bm k)|^2 - \frac{2}{3} |\varphi(\bm k')|^2\right]\varphi(\bm k')}_{\mbox{\tiny electron-hole attraction} \propto u_{\bm k'} v_{\bm k'} (u_{\bm k}^2 - |v_{\bm k}|^2)} \\
+ \underbrace{ V_{\bm k - \bm k'} 2\varphi(\bm k)|\varphi(\bm k')|^2}_{\mbox{\tiny e-e and e-h repulsion}}  .
\end{multline}
%\end{widetext}
These $\mathcal M$ contribution leads to the exciton instability.
%In derivations we took into account that 
%\begin{subequations}
%\label{useful}
%\begin{align}
%S^\dag a_{\bm k}^\dag a_{\bm k} S = u_{\bm k}^2 a_{\bm k}^\dag a_{\bm k} + |v_{\bm k}|^2 b_{-\bm k} b_{-\bm k}^\dag + u_{\bm k} v_{\bm k} a_{\bm k}^\dag b_{-\bm k}^\dag + u_{\bm k} v_{\bm k}^* b_{-\bm k} a_{\bm k},\label{a}\\
%S^\dag b_{-\bm k}^\dag b_{-\bm k} S = u_{\bm k}^2 b_{-\bm k}^\dag b_{-\bm k} + |v_{\bm k}|^2 a_{\bm k} a_{\bm k}^\dag - u_{\bm k} v_{\bm k} b_{-\bm k}^\dag a_{\bm k}^\dag - u_{\bm k} v_{\bm k}^* a_{\bm k} b_{-\bm k},\label{b}\\
%S^\dag a_{\bm k}^\dag a_{\bm k}^\dag S = u_{\bm k}^2 a_{\bm k}^\dag a_{\bm k}^\dag + {v_{\bm k}^*}^2 b_{-\bm k} b_{-\bm k} + u_{\bm k} v_{\bm k}^*(a_{\bm k}^\dag b_{-\bm k} + b_{-\bm k} a_{\bm k}^\dag),\\
%S^\dag b_{-\bm k}^\dag b_{-\bm k}^\dag S = u_{\bm k}^2 b_{-\bm k}^\dag b_{-\bm k}^\dag + v_{\bm k}^2 a_{\bm k} a_{\bm k} + u_{\bm k} v_{\bm k}(a_{\bm k}b_{-\bm k}^\dag  + b_{-\bm k}^\dag a_{\bm k}),\\
%S^\dag a_{\bm k}^\dag b_{-\bm k}^\dag S = u_{\bm k}^2 a_{\bm k}^\dag b_{-\bm k}^\dag - {v_{\bm k}^*}^2 b_{-\bm k} a_{\bm k} + u_{\bm k} v_{\bm k}^*(b_{-\bm k} b_{-\bm k}^\dag  - a_{\bm k}^\dag a_{\bm k}),\label{e}\\
%S^\dag b_{-\bm k} a_{\bm k} S = u_{\bm k}^2 b_{-\bm k}a_{\bm k} - v_{\bm k}^2  a_{\bm k}^\dag b_{-\bm k}^\dag + u_{\bm k} v_{\bm k}(b_{-\bm k} b_{-\bm k}^\dag  - a_{\bm k}^\dag a_{\bm k}).\label{f}
%\end{align}
%\end{subequations}

The order parameter can be found from the condition that, in the ground state, the contributions that give rise to electrons and holes vanish.\cite{GuseinovKeldysh,keldysh68a,tolmachev58} Neglecting exciton-exciton interactions we have the condition of vanishing ``dangerous'' terms in the form
\begin{equation}
\label{F=0}
\lambda_{\bm k} + M_{\bm k} =0.
\end{equation}

In this case of absence interband coupling, $\lambda_{\bm k}=0$, naturally, one solution of homogeneous Eq.~\eqref{F=0} corresponds to $\varphi(\bm k)\equiv 0$ (no order parameter). A non-trivial solution is possible and can be found iteratively, assuming that $|\varphi(\bm k)| \ll 1$. 

\emph{$1^{\rm st}$ approximation:} we keep only $\varphi(\bm k)$-linear terms in~\eqref{Mk} and obtain
\begin{equation}
\label{order0}
0 = \left[E_e(\bm k) + E_h({-}\bm k)\right]\varphi(\bm k) - \sum_{\bm k'} V_{\bm k - \bm k'} \varphi(\bm k'),
\end{equation}
or
\begin{equation}
\label{order0:1}
0 = \left[E_g + \frac{\hbar^2 k^2}{2\mu}\right]\varphi(\bm k) - \sum_{\bm k'} V_{\bm k - \bm k'} \varphi(\bm k'),
\end{equation}
where $\mu= m_e m_h/(m_e+m_h)$ is the electron-hole reduced mass. It can be assumed that $\varphi(\bm k) = a\varphi^{1s}_{\bm k}$, where $\varphi^{1s}_{\bm k}$ is the (normalized) exciton wavefunction corresponding to the lowest in the energy bound state with the energy $E_0=-E_B$ where $E_B>0$ is the exciton binding energy:
\begin{equation}
\label{bound}
\frac{\hbar^2 k^2}{2\mu}\varphi^{1s}_{\bm k}- \sum_{\bm k'} V_{\bm k - \bm k'} \varphi^{1s}_{\bm k} = - E_B \varphi^{1s}_{\bm k}.
\end{equation} 
In this way, for $E_g=E_B$, Eq.~\eqref{order0:1} has nontrivial solution.

\emph{$2^{\rm nd}$ approximation:} We next show that for $E_g < E_B$ the nontrivial solution is present as well. Keeping cubic terms we obtain
\begin{equation}
\label{order3}
\left[E_g + \frac{\hbar^2 k^2}{2\mu}\right]\varphi(\bm k) - \sum_{\bm k'} V_{\bm k - \bm k'} \varphi(\bm k') = \mathcal F^{(3)}_{\bm k}\{\varphi\},
\end{equation}
where the functional 
\begin{multline}
\label{F3}
\mathcal F^{(3)}_{\bm k}\{\varphi\}= \frac{2}{3}\left[E_g + \frac{\hbar^2 k^2}{2\mu}\right]|\varphi(\bm k)|^2\varphi(\bm k) \\
+ \sum_{\bm k'} V_{\bm k - \bm k'} \left[2\varphi(\bm k) \varphi^*(\bm k')-2|\varphi(\bm k)|^2-\frac{2}{3}|\varphi(\bm k')|^2 \right]\varphi(\bm k').
\end{multline}
It is convenient to simplify these equations making formal replacement 
\begin{equation}
\label{replace}
\varphi(\bm k) \to \varphi(\bm k)\left[1-\frac{2}{3}|\varphi(\bm k)|^2 \right].
\end{equation}
Equation~\eqref{order3} remains, while the functional takes a simpler form
\begin{equation}
\mathcal F^{(3)}_{\bm k}\{\varphi\}= 2\sum_{\bm k'} V_{\bm k - \bm k'} \varphi(\bm k)\left[ \varphi^*(\bm k')-\varphi^*(\bm k)\right]\varphi(\bm k').
\end{equation}

Formal solution of Eq.~\eqref{order3} can be written via the Greens function of the exciton $G_{exc}(\bm k;\varepsilon)$:
\begin{equation}
\label{formal:sol}
\varphi(\bm k) = - \sum_{\bm k'} G_{exc}(\bm k-\bm p;-\Delta) \mathcal F^{(3)}_{\bm p}, 
\end{equation}
where
\begin{multline}
\label{exc:Greens}
\left[-\varepsilon + \frac{\hbar^2 k^2}{2\mu}\right]G_{exc}(\bm k-\bm p;\varepsilon)- \sum_{\bm k'} V_{\bm k - \bm k'}G_{exc}(\bm k'-\bm p;\varepsilon)\\
= -\delta_{\bm k,\bm p}.
\end{multline} 

To find the ground state, we  set 
\begin{equation}
\label{varphi}
\varphi(\bm k) = |a| e^{\mathrm i \chi} \varphi_{\bm k}^{1s},
\end{equation}
 and take into account the cubic-in-$\varphi$ contributions in $M_{\bm k}$. For $M_{\bm k}$ in the form of Eq.~\eqref{Mk} it turns out that the phase $\chi$ is arbitrary, while $a$ can be found from the non-linear equation
\begin{equation}
\label{order3:1}
0=\sum_{\bm k} \left(\varphi_{\bm k}^{1s}\right)^* M_{\bm k} \quad \Rightarrow \quad \left(E_g - E_B\right)|a| = -g|a|^3,
\end{equation}
where the constant $g$
\begin{multline}
\label{f:F3}
g= 2\sum_{\bm k} \left(\varphi_{\bm k}^{1s}\right)^*\mathcal F_3^{(3)}\{\varphi_{\bm k}^{1s}\} \\
= \sum_{ \bm k} \left(E_B + \frac{\hbar^2 k^2}{2\mu}\right)|\varphi_{\bm k}^{1s}|^4 - 2\sum_{\bm k,\bm k'} V_{\bm k - \bm k'} |\varphi_{\bm k}^{1s}|^2|\varphi_{\bm k'}^{1s}|^2.
\end{multline}
In derivation of Eq.~\eqref{f:F3} we made use of the fact that function $\varphi_{\bm k}^{1s}$ obeys Eq.~\eqref{bound}. One can check that $g>0$ and show that an order of magnitude estimate gives  $g\sim a_B^2 E_B$. Equations~\eqref{order3:1} and \eqref{f:F3} agree with Ref.~\onlinecite{GuseinovKeldysh}. The result presented in the main text corresponds to the first term in Eq.~\eqref{f:F3}.

\section{Exciton dispersion in the presence of light-matter interaction}\label{sec:exc:LT:standard}

It is instructive to establish a link between the results obtained in the main text and well-known effect of the exciton longitudinal-transverse splitting in two-dimensional systems.\cite{ivchenko05a,PhysRevB.41.7536,maialle93,goupalov98,Yu:2014fk-1,glazov2014exciton} We start from a non-condensed isotropic case. In this situation it is convenient to use the basis of the transversal and longitudinal exciton modes with $\bm P \parallel \bm Q$ and $\bm P \perp \bm Q$, respectively. The polarizability of any of the modes is given by Eq.~\eqref{pi1:normal}, while the Greens function for the electric field can be recast as [cf.~\eqref{greens:EE}]
\begin{subequations}
    \label{D:E:LT}
    \begin{equation}
       \label{D:E:L}
       D_L^E(\Omega, \bm Q) = -2\pi\sqrt{Q^2 - (\Omega/c)^2 -\mathrm i0},
    \end{equation}
\begin{equation}
    \label{D:E:T}
    D_T^E(\Omega, \bm Q) = \frac{2\pi(\Omega/c)^2}{\sqrt{Q^2 - (\Omega/c)^2 -\mathrm i0}}.
\end{equation}    
\end{subequations}
The dispersion of excitons with allowance for the light-matter interaction can be found from the self-consistency requirement with the result:
\begin{equation}
    \label{exc:LT:self}
    \pi_1(\Omega,\bm Q) D_{L,T}^E(\Omega, \bm Q)=1.
\end{equation}
At $\bm Q=0$ we obtain the radiative damping of the excitons in the form
\begin{equation}
    \label{Gamma:0}
    \hbar\Gamma_0 = 2\pi  \frac{E_g - E_B}{\hbar c}d_{\rm exc}^2.
\end{equation}
The exciton LT-splitting (at $Q\gg(E_g-E_B)/\hbar c$, i.e., where the retardation is not important) is
\begin{equation}
    \label{LTspltting}
    \Delta E_{LT} = 2\pi Q d_{\rm exc}^2 = \frac{\hbar c Q}{E_g-E_B} \hbar\Gamma_0,
\end{equation}
in full agreement with previous works. 

Interestingly, for non-condensed excitons with $E_B = E_g$ (just at the threshold of exciton insulator instability) the dispersion of the longitudinal and transverse branches as $Q\to 0$ is given by
\begin{subequations}
\begin{equation}
\label{E:0:L}
    E_L^0(Q) = \frac{d_{\rm exc}^2 Q}{1+ \frac{d_{\rm exc}^4}{\hbar^2 c^2}} \approx \begin{cases}
    d_{\rm exc}^2 Q, d_{\rm exc}^2 \ll \hbar c,\\
    \frac{\hbar^2 c^2}{d_{\rm exc}^4} Q, \quad d_{\rm exc}^2 \gg \hbar c.
    \end{cases}
\end{equation}
\begin{equation}
\label{E:0:T}
    E_T^0(Q) = \frac{\hbar^2 Q^2}{2M} - 2\pi d_{\rm exc}^2\frac{\hbar^2 Q^3}{(2Mc)^2}.
\end{equation}
\end{subequations}
While the transversal branch remains parabolic with small correction $\propto Q^3$, the longitudinal branch acquires a linear dispersion. Its propagation velocity is always smaller that $c$ because of the causuality imposed by the light-matter interaction.

\section{Evaluation of $N_{\rm exc}$}

According to Eq.~\eqref{N:exc:gen} we have in the limit of very low temperatures where the small-$Q$, small-$\varphi$ asymptotics is valid
\begin{equation}
    \label{N:exc:conv:1}
    N_{\rm exc} = {\frac{k_B T}{2(2\pi)^3 d_{\rm exc}^2}\int \frac{Q^2 dQ d\varphi}{Q^2{\varphi}^2+(\xi/c)^2 Q^3}},
\end{equation}
where the limits of the integral should be chosen in such a way that $\mathcal E(\bm Q)\lesssim k_B T$. Using the interpolation  expression for the energy~\eqref{disper:interp} we obtain $|\varphi|<\varphi_Q \equiv k_B T/(\hbar cQ)$ and the angular integral is evaluated as
\[
\int_{-\varphi_Q}^{\varphi_Q} d\varphi \frac{Q^2}{Q^2{\varphi}^2+(\xi/c)^2 Q^3} =\frac{2c}{\sqrt{Q} \xi}\tan^{-1}{\left(\frac{k_B T}{\hbar \xi Q^{3/2}}\right)}.
\]
The remaining integral over $Q$ can now be easily calculated giving the result
\begin{equation}
    N_{\rm exc} = \frac{k_B T}{2(2\pi)^3 d_{\rm exc}^2} \times  \frac{4\pi c}{\sqrt{3}} \left(\frac{ k_B T}{\hbar\xi^4} \right)^{1/3},
\end{equation}
in agreement with Eq.~\eqref{N:exc:conv} of the main text.

%\bibliography{all-1}
%\bibliography{AI}
%\bibliography{AI1}


\end{document}